\documentstyle[12pt,epsf]{article}

\newcommand{\beq}{\begin{equation}}
\newcommand{\eeq}{\end{equation}}
\newcommand{\beqa}{\begin{eqnarray}}
\newcommand{\eeqa}{\end{eqnarray}}
\newcommand{\ba}{\begin{array}}
\newcommand{\ea}{\end{array}}
\newcommand{\CR}{\nonumber \\}
\newcommand{\pa}{\partial}

\newcommand{\B}{\beta}

\newcommand{\half}{{1\over 2}}
\newcommand{\Tr}{{\rm Tr}}

\newcommand{\eq}{\begin{equation}}
\newcommand{\en}{\end{equation}}
\newcommand{\eqn}{\begin{eqnarray}}
\newcommand{\enn}{\end{eqnarray}}

\newcommand{\cA}{{\cal A}}
\newcommand{\cH}{{\cal H}}
\newcommand{\norm}[1]{\parallel #1\parallel}
\newcommand{\ol}{\overline}
\newcommand{\ch}{ {\rm ch} }
\newcommand{\Td}{ {\rm Td} }
\newcommand{\Index}{ {\rm Index}\,}
\newcommand{\diag}{ {\rm diag} }
\newcommand{\AD}[1]{$\overline{\mbox{D~\,}}\!\!\!$#1}
\newcommand{\Dbar}{$\overline{\mbox{D}}$}
\newcommand{\bra}[1]{\left\langle\, #1\,\right|}
\newcommand{\ket}[1]{\left|\, #1\,\right\rangle}
\newcommand{\wt}{\widetilde}
\newcommand{\wh}{\widehat}
\newcommand{\del}{\partial}
\newcommand{\sq}{\sqrt{2}}
\newcommand{\lra}{\leftrightarrow}

\makeatletter
\@addtoreset{equation}{section}
\@addtoreset{footnote}{page}
\makeatother

\def\mat#1{\matt[#1]}
\def\matt[#1,#2,#3,#4]{\left(%
\begin{array}{cc} #1 & #2 \\ #3 & #4 \end{array} \right)}

\def\Z {{\bf Z}}
\def\Q {{\bf Q}}
\def\R {{\bf R}}
\def\C {{\bf C}}
\def\B {{\bf B}}
\def\K {{\bf K}}

\begin{document}

\makeatletter
\def\setcaption#1{\def\@captype{#1}}
\makeatother

\begin{titlepage}
\vspace*{-2.5cm}
\null
\begin{flushright}
hep-th/0108085  \\
YITP-01-54\\
CITUSC/01-029\\
UT-958\\
August, 2001
\end{flushright}
\vspace{0.5cm}
\begin{center}
{\LARGE \bf
D-branes, Matrix Theory and K-homology
\par}
\lineskip .75em
\vskip1.5cm
\normalsize

{\large
 Tsuguhiko Asakawa\footnote{
E-mail:\ \ asakawa@yukawa.kyoto-u.ac.jp    },
 Shigeki Sugimoto\footnote{
E-mail:\ \ sugimoto@citusc.usc.edu}  and
Seiji Terashima\footnote{
E-mail:\ \ seiji@hep-th.phys.s.u-tokyo.ac.jp} }
\vskip 2.5em
{
${}^1$
\it
Yukawa Institute for Theoretical Physics,
Kyoto University,\\
Kyoto 606-8502, Japan
}
\vskip 1em
{
${}^2$
 \it
CIT/USC Center for Theoretical Physics,
University of Southern California,\\
 Los Angeles CA90089-2535, USA \\}
\vskip 1em
{
${}^3$
 \it  Department of Physics, Faculty of Science,
 University of Tokyo,\\
Tokyo 113-0033, Japan}
\vskip 2em

{\bf Abstract}
\end{center} \par

In this paper, we study a new matrix theory
 based on
non-BPS D-instantons in type IIA string theory and
D-instanton - anti D-instanton system in type IIB string theory,
which we call K-matrix theory.
The theory correctly incorporates the creation and annihilation
processes of D-branes.
The configurations of the theory are identified with spectral triples,
which are the noncommutative generalization of
 Riemannian geometry \'a la Connes,
and they represent the geometry on the world-volume of
higher dimensional D-branes.
Remarkably, the configurations of D-branes in the K-matrix theory
are naturally classified by a K-theoretical version of homology group,
called K-homology.
Furthermore, we argue that the K-homology correctly
classifies the D-brane configurations
from a geometrical point of view.
We also construct the boundary states corresponding
to the configurations of the K-matrix theory,
and explicitly show that they represent
the higher dimensional D-branes.

\newpage

\tableofcontents

\end{titlepage}

\baselineskip=0.7cm

\section{Introduction}
\setcounter{equation}{0}

The recent development of non-BPS systems shows
that it is essential to incorporate the creation and
annihilation process of non-BPS D-branes or
D-brane - anti D-brane pairs in string theory.
For example,
as shown in \cite{DiMoWi},
there are some examples that
D-branes wrapped on non-trivial cycles
decay through the creation and annihilation process.
In fact, taking these process into account,
D-brane charges are successfully classified by K-theory
\cite{Wi}, which shows that
(co)homology groups are no longer basic tools for
the classification of topologically stable configurations
of D-branes.
It also suggests that the usual description of RR-fields as
differential forms is, in general, insufficient to correctly
describe the background of string theory \cite{MoWi}.
Therefore, it seems to be very important to find a
non-perturbative formulation of string theory, in which
the creation and annihilation process of
D-branes is correctly incorporated, as in the second
quantization of field theory.

The K-theory structure of the theory  appears most naturally
in the world-volume gauge theory of space-time filling brane systems,
i.e.
 non-BPS D9-brane system in type IIA string theory \cite{Ho}
 and D9-\AD9 system in type IIB string theory \cite{Wi}.
However, since the ten dimensional gauge theories are non-renormalizable,
it is hard to consider this kind of theory as a fundamental
theory. So, lower dimensional systems are preferable for our purpose.
One of the most interesting possibilities is the lowest dimensional case,
namely matrix theory.
The matrix formulations of
type II string theory or M-theory are proposed in
\cite{BFSS,IKKT,DiVeVe} etc.
However, the K-theory structure is not clear in the framework
of these matrix theories.
One of the reasons is that
the matrix model of \cite{BFSS}, for example,
is formulated as the theory of infinite number
of D-particles, which corresponds to
the infinite momentum frame in M-theory
and there are no anti D-particles.
Similarly, K-theory structure of the IIB matrix
theory \cite{IKKT}
should be realized in a completely different
way from the D9-\AD9 system since
it is not based on the
D-\Dbar~system.

In this paper, we propose a new type of matrix theory based on
non-BPS D-instantons in type IIA theory and
D-instanton - anti D-instanton system in type IIB theory.
If one prefer Minkowski space-time rather than Euclidean one,
it could be more interesting to consider matrix theories
based on
D-particle - anti D-particle system in type IIA theory and
non-BPS D-particles in type IIB theory.
But, since the formulation of the latter
is quite analogous to the former,
we mainly deal with the former for simplicity.

It is tempting to consider such matrix theory
as the fundamental theory of covariant type II
string theory.
Unfortunately, however, we do not know
the precise form of the action of the theory.
We leave
the attempt to give a precise formulation
and the argument about the consistency of the theory
as future problems.
Nevertheless, as we will see in the following sections,
it is possible to analyze
the topologically stable configurations
of the theory.
The configurations of the theory can be interpreted as
the higher dimensional D-branes, and we will show
in section \ref{K-homo} that they are classified by
a K-homology group, which is a dual of K-theory group.
Therefore, the theory
successfully recovers the K-theory structure.
For this reason, we would like to refer to these matrix
theories as K-matrix theory.

At first sight,
it seems to be strange that the K-homology naturally
appears in the K-matrix theory instead of the K-theory.
However, it turns out that
the configurations of the K-matrix theory
represent the world-volume of D-branes,
which should be classified by some kinds of
homology theory.
The K-homology carries not only the information
of cycles but also that of gauge bundles on them.
Hence the K-homology is a
natural candidate for the classification
of D-brane configurations.

In the formulation of matrix theories,
geometry does not play an important role,
since the world-volume manifold is just a set of zero dimensional
points. There are no non-trivial topology, metric, gauge bundle,
and so on.
Instead, the algebra among the matrix variables
becomes important.
In the K-matrix theory, the size of the matrix variables
are infinity from the beginning, so that arbitrary numbers
of non-BPS D-instantons or D-instanton - anti D-instanton pairs
can be created. Therefore, the matrices should be thought of as
linear operators acting on an infinite dimensional Hilbert space.
One of the interesting features in the K-matrix theory is that
the geometric information is hidden in this operator algebra.
In fact, it is well-known that
there is a one-to-one correspondence between
operator algebras and topological spaces.
More precisely,
 the *-isomorphism classes of
commutative $C^*$-algebras
are in one-to-one correspondence with
  the homeomorphism classes
of locally compact Hausdorff spaces.
Here, a $C^*$-algebra is a norm closed self-adjoint subalgebra of
the bounded operator algebra $\B(\cH)$ for some Hilbert space $\cH$.
This fact is the starting point of the noncommutative geometry.
Remarkably, the variables in the K-matrix theory
have a direct interpretation in terms of the noncommutative geometry.
We will argue in section \ref{SpecTriple}
that these variables define a spectral triple,
which is introduced by Connes to define a
noncommutative generalization of the Riemannian geometry
\cite{Connes},
and explain that the spectral triples represent
the configurations of D-branes, that is, the geometry represented
by a spectral triple is nothing but the world-volume geometry
of the corresponding D-branes.

The K-homology is defined as
equivalence classes of the spectral triples
under some equivalence relations.
The equivalence relations also have physical interpretations,
that is, continuous deformation
(homotopy equivalence), gauge equivalence and
 creation and annihilation of non-BPS D-instantons
or D-instanton - anti D-instanton pairs.
These observations suggest
that the D-brane configurations
in the K-matrix theory are classified by the K-homology groups.
We also have a topological description of K-homology,
which enables us to interpret the classification in terms
of geometry of the D-brane world-volume.
We will explain these facts in detail in section
\ref{K-homo}.

We also analyze these facts using boundary state approach
of D-branes along the line with the work of \cite{Is},
in which boundary states of  higher dimensional D-branes
are constructed from the boundary states of
noncommutative configurations of D-instantons in bosonic string theory.
We can formally construct a boundary state
corresponding to each configuration of the K-matrix theory.
One of the merits of this approach is that
we can avoid the ambiguity in
the action of the K-matrix theory.
Actually,
this approach is closely related to the BSFT approach
and we can implicitly use the BSFT action by analyzing the
boundary states.
Thanks to this fact, as we will see in section \ref{BdryState},
we can explicitly show that a canonical choice of
the spectral triples
represents a higher dimensional D-brane.
In other words, we can explicitly see how the operators
in the K-matrix theory represent the geometry of D-branes
using string theory.
This gives another viewpoint for
 our proposal that D-branes are represented by spectral
triples, and also for the correspondence between operator
algebras and topological spaces.

The paper is organized as follows.
In section \ref{K-matrix}, we outline
the basic structure of the K-matrix theory,
and explain how the configurations with finite action
can be expressed.
 Section \ref{SpecTriple} deals with the geometric
interpretation of the configurations. We explain
that they are related to the spectral triples
and interpreted as the configurations of
higher dimensional D-branes.
In section \ref{K-homo}, we claim that
the D-brane configurations in the K-matrix theory
are classified by K-homology.
Some comments on the generalization to
the higher dimensional cases are made in section \ref{KK-theory}
from the viewpoint of KK-theory.
We also discuss
that the Chern-Simons terms are given by the
Chern character of the K-homology.
Section \ref{BdryState} is devoted to
the boundary state approach.
We will construct the boundary states
of higher dimensional D-branes
from the boundary state of non-BPS D-instantons
with tachyon condensation.
Note also that the calculation performed in this section
provides an efficient way to obtain the
correct tensions and RR-charges of D-branes
made via tachyon condensation.
Finally,
we make some speculative discussions in section \ref{Discuss}.

\section{K-matrix theory}
\label{K-matrix}
\subsection{Type IIA K-matrix theory}
\label{IIA}

In type IIA string theory, the lowest dimensional D-brane
is the non-BPS D-instanton.
In order to obtain the exact action
of the theory explicitly,
we may need powerful
symmetries such as supersymmetry, conformal symmetry, etc.
The nonlinearly realized supersymmetry with 32 supercharges,
which is based on the idea that the vacuum with a non-BPS D-brane belongs
to the spontaneously broken phase of the supersymmetry
\cite{Sen,Yo},
may be strong enough to determine the action, as calculated
in \cite{TeUe} up to some order.
Up to now, however,
we do not know how to write down the action of the theory exactly.
Nevertheless, as we will see,
we can extract some topological information
of the theory, on which we mainly focus in this paper,
without knowing the detailed structure of the theory.

Let us summarize
the main ingredients of the non-BPS D-instanton
theory.
The gauge group of the theory
with $N$ non-BPS D-instantons is $U(N)$. The bosonic
part consists of ten scalar fields $\Phi^\mu$ ($\mu=0,\dots,9$) and
a tachyon $T$.
They are self-adjoint (Hermitian) matrices and
 belong to the adjoint representation of the gauge group.
\footnote{
Throughout this paper, we only consider the tachyons and massless
modes. Massive modes are considered to be integrated out or
neglected.}

The important point is that in order to incorporate
creation and annihilation of the non-BPS D-instantons,
we must take the limit $N=\infty$, so that
arbitrary numbers of non-BPS D-instantons can be created.
Therefore, the vector space, on which matrices $\Phi^\mu$ and $T$ are
represented, is an infinite dimensional vector space.
We assume that this vector space is a Hilbert space $\cH$.
This Hilbert space should be separable,
i.e. it has countably many orthonormal basis, since
there is a  one-to-one correspondence between
the basis of $\cH$ and the Chan-Paton
indices of non-BPS D-instanton.
Note that since every infinite dimensional separable Hilbert space is
isomorphic to $l^2({\bf N})$, we can uniquely associate
the Hilbert space $\cH$
with the Chan-Paton indices of the non-BPS D-instantons up to isomorphism.
Then $\Phi^\mu$ and $T$ are regarded as linear operators acting on the
Hilbert space $\cH$.

The action is basically obtained through the dimensional reduction
of the non-BPS D9-brane action.
 The kinetic terms and the tachyon potential
are roughly given as
\begin{eqnarray}
S(T,\Phi^\mu)
&\sim&\Tr\left(e^{-T^2}
[\Phi^\mu,\Phi^\nu]^2 + e^{-T^2}[\Phi^\mu,T]^2 + e^{-T^2}+\cdots
\right).
\label{action}
\end{eqnarray}
Actually, using
the boundary string field theory \cite{KuMaMo2,Ts2,KrLa,TaTeUe},
the action for non-BPS D-instantons \cite{TeUe,An,KuMaMo2} is
calculated as
\beqa
\label{BSFTaction}
S
&=&-\sqrt{2} T_{-1}
\Tr  \left( e^{-\frac{1}{4}T^2} \sqrt{\det(\delta_{\mu\nu}
-i [ \Phi_{\mu} , \Phi_{\nu}])} \, {\mathcal F}
\left[\frac{1}{4\pi} {\mathcal G}^{\mu\nu}
[ \Phi_{\mu} , T] [ \Phi_{\nu} , T]
\right] \right),
\eeqa
where
${\mathcal G}^{\mu\nu}\equiv
\left(\frac{1}{1 -i [ \Phi_{\mu} , \Phi_{\nu}]}\right)^{(\mu\nu)}$ and
${\mathcal F}[x]\equiv\frac{4^x x (\Gamma(x))^2}{2\Gamma(2x)}
=1+2(\log 2)x+O(x^2)$.
Here, $(\mu\nu)$ indicates the symmetrization.
Though the action is not exact, i.e.
the higher commutator terms are neglected,
the action (\ref{BSFTaction}) may be
a rather good starting point to consider the non-BPS D-branes.
In fact, the action successfully describes the tachyon condensation
and contains the Dirac-Born-Inferd type action.

{}From this it is reasonable to expect that
the full action has the form (\ref{action})
and then the action is roughly estimated by the following inequality.
\begin{eqnarray}
|S(T,\Phi^\mu)|
&\leq&\Tr\, e^{-T^2}\left(
\norm{[\Phi^\mu,\Phi^\nu]}^2 +\norm{[\Phi^\mu,T]}^2 + 1\,
\right) +\cdots,
\label{action2}
\end{eqnarray}
where $\norm{\cdot}$ is the operator norm.
In order to obtain configurations with finite action,
it seems to be natural to require
\begin{eqnarray}
\Tr\,e^{-T^2}<\infty,~~
\norm{[\Phi^\mu,\Phi^\nu]}<\infty,~~\norm{[\Phi^\mu, T]}<\infty,
\label{finiteness}
\end{eqnarray}
that is,
 $[\Phi^\mu,\Phi^\nu]$ and  $[\Phi^\mu, T]$
are bounded operators for $\mu,\nu=0,1,\dots,9$,
and  $e^{-T^2}$ is traceclass.

In particular, the first condition in
(\ref{finiteness}) implies that
 the tachyon $T$ is not a bounded operator.
To see this fact,
let $\{\lambda_n \}$ be the set of eigenvalues of the operator $T$.
In order for the potential $\Tr\,e^{-T^2}$ to be finite,
each eigenvalue should have finite multiplicity, and
$|\lambda_n|\rightarrow\infty$ as $n\rightarrow\infty$.
Since the norm of an operator is larger than or equal to
its largest eigenvalue, the norm of $T$ diverges to infinity.

To avoid this infinity, it may be convenient to
use the bounded operator
\begin{eqnarray}
T_b=\frac{T}{\sqrt{1+T^2}}
\end{eqnarray}
normalized such that $T_b^2=1$ is the minimum of the potential.
In this normalization,
the eigenvalues of $T_b^2$ accumulates to $1$
in order to obtain a finite energy configuration.
In other words, we require $T_b^2-1$ to be a compact operator.
Here an operator $K$ on $\cH$ is said to be compact
if it has an expansion
\begin{eqnarray}
K=\sum_{n\ge0}\mu_n
\ket{\psi_n}\bra{\phi_n},
\end{eqnarray}
where, $\mu_n\rightarrow0$ as $n\rightarrow\infty$,
 $\{\psi_n\}_{n\in {\bf N}}$ and $\{\phi_n\}_{n\in {\bf N}}$
are orthonormal sets.

Similarly, the requirement that
$[\Phi^\mu,T]$ ($\mu=0,1,\dots,9$) are  bounded operators
implies that $[\Phi^\mu,T_b]$ are compact operators.
This condition is analogous to the condition
$D_\mu T(x)\rightarrow0$ at infinity in usual field theory.
Therefore we should, at least,
 require the self-adjoint bounded operator $T_b$ to
satisfy the conditions
\begin{eqnarray}
T_b^2-1\in \K(\cH),~~~[\Phi^\mu,T_b]\in \K(\cH),
~~~(\mu=0,1,\dots,9),
\label{cptness}
\end{eqnarray}
where
$\K(\cH)$ denotes the set of compact operators
on $\cH$.

Strictly speaking, we don't know whether these conditions for
the operators $\Phi^\mu$ and $T$ are necessary nor sufficient
ones for the finiteness
of the action, since we don't know the exact action
of the theory.
In the following sections, we will see that our proposal is
highly plausible. They beautifully fit the mathematical
framework of noncommutative geometry
and K-homology.
We will also examine some examples and see how they work.

\subsection{Type IIB K-matrix theory}
\label{IIB}
The argument in the previous subsection
can also be applied to type IIB string theory.
There are BPS D-instantons in type IIB string theory, and
the matrix theory related to the D-instantons is constructed
 in \cite{IKKT}. In order to incorporate creation and annihilation
of D-instanton - anti D-instanton pairs in the theory,
we should add degrees of freedom of the anti D-instantons.
Thus the matrix theory we consider here is based on
the D-instanton - anti D-instanton system.
The world-point theory of $N$ D-instantons and $M$ anti D-instantons
has $U(N)\times U(M)$ gauge symmetry. There are
ten pairs of scalar fields $\Phi^\mu$, $\ol\Phi^\mu$ ($\mu=0,\dots,9$) and
a tachyon field $T$. $\Phi^\mu$, $\ol\Phi^\mu$ and $T$
are in (adjoint,1), (1,adjoint) and $(N,M)$
 representation of the gauge group $U(N)\times U(M)$, respectively.

We take both $N$ and $M$ to be infinity.
The Chan-Paton Hilbert space should be $\Z_2$-graded as
$\wh\cH=\cH^{(0)}\oplus\cH^{(1)}$, where
the basis of $\cH^{(0)}$ and $\cH^{(1)}$ is in
one-to-one correspondence with the D-instantons
and the anti D-instantons, respectively.
The scalar fields $\Phi^\mu$ and $\ol\Phi^\mu$
are operators acting on $\cH^{(0)}$ and
$\cH^{(1)}$, respectively, while the tachyon
$T$ is an operator from $\cH^{(0)}$ to  $\cH^{(1)}$.

As discussed in the previous subsection,
there are some constraints on these operators,
which are analogous to (\ref{finiteness}) and
(\ref{cptness}) for the IIA K-matrix theory.
If we use the normalized tachyon field $T_b$,
such that the minimum of the potential is given by
$T_b^*T_b=T_b T_b^*=1$,
the same argument as in (\ref{cptness})
implies that $T_b$ should be an element of
$\B(\cH^{(0)},\cH^{(1)})$,
bounded linear operators from $\cH^{(0)}$ to  $\cH^{(1)}$,
 which satisfies
\begin{eqnarray}
&&T_b^*T_b-1\in\K(\cH^{(0)}),
~~~T_bT_b^*-1\in\K(\cH^{(1)}),
\label{Pot}
\\
&&T_b\Phi^\mu-\ol\Phi^\mu T_b\in\K(\cH^{(0)},\cH^{(1)}),
~~~(\mu=0,1,\dots,9).
\label{Tkin}
\end{eqnarray}
Here
$\K(\cH^{(0)},\cH^{(1)})$ denotes the set of compact operators
from $\cH^{(0)}$ to $\cH^{(1)}$.
The conditions (\ref{Pot}) and (\ref{Tkin}) are
again expected from the finiteness of the potential
and kinetic terms of the tachyon, respectively.

We can rewrite these conditions just like
(\ref{cptness}) as
\begin{eqnarray}
\wh F^2-1\in \K(\wh\cH),~~~[\wh\Phi^\mu,\wh F]\in \K(\wh\cH),
~~~(\mu=0,1,\dots,9),
\end{eqnarray}
where $\wh\Phi^\mu=\diag(\Phi^\mu,\ol\Phi^\mu)$ and
$\wh F=\left({~~~T_b^*\atop T_b~~~} \right)$.

Note that there are certain similarities
 between this IIB K-matrix theory
and the model of \cite{CoLo} which is based on
the gauge theory on two points.
Actually, the D-\Dbar~pair in the K-matrix theory
 represents the discrete two points
and the tachyon $T$ corresponds to the Higgs field in \cite{CoLo,Connes},
though the precise action is not exactly the same.

\subsection{Chern-Simons terms and D-brane configurations}
\label{CSandD}

Although we don't have the exact action for the K-matrix theory,
we can use some exact results from the calculation in
 BSFT \cite{KuMaMo2,KrLa,TaTeUe}. In particular,
Chern-Simons terms are known exactly,
at least in the case that the RR-fields are constant.
(See \cite{Wy,KrLa} for the comments on the corrections.)

Let us briefly review the Chern-Simons terms in
the action  of $N$ non-BPS D-instantons in type IIA string theory.
First we introduce fermions $\psi_1^\mu,\psi_2^\mu$ ($\mu=0,\dots,9$),
which represent $SO(10,10)$ gamma matrices,
satisfying the anti-commutation relations
\beq
\{ \psi_1^\mu, \psi_2^\nu\}=\delta^{\mu\nu}
,\;\; \{ \psi_1^\mu, \psi_1^\nu\}=\{ \psi_2^\mu, \psi_2^\nu\}=0.
\label{1010}
\eeq

Then the Chern-Simons term obtained in  \cite{TaTeUe}
can be written as
\begin{eqnarray}
S_{\rm CS}
&=&{\rm Sym} \Tr_N \Tr_2 \left( \sigma^1  \Tr_{\psi}
\left( \wh{C} e^{Z^2} \right)\right)\\
&=& {\rm Sym} \Tr_N \Tr_2 \left( \sigma^1  \Tr_{\psi} \left( \wh{C}
e^{\left(-T^2+\frac{1}{2}[\Phi^\mu,\Phi^\nu]\psi_2^\mu\psi_2^\nu
+i[\Phi^\mu,T]\psi_2^\mu\sigma^1\right)}
\right)\right),
\label{CSterm}
\end{eqnarray}
where
\beqa
iZ &=& -i \Phi^\mu \psi_2^\mu+ T \sigma^1,
\label{Z}\\
\wh{C} &=& \sum_n C_{\mu_1\cdots\mu_n}(\Phi)
\, \psi_1^{\mu_1} \cdots \psi_1^{\mu_n}.
\label{C}
\eeqa
Note that $\sigma^1=\left({~~1\atop1~~}\right)$ also behaves
as a fermion.

Here $Z$ is regarded as a $2\times 2$ matrix,
whose components are also matrices,
and $\Tr_2$ denotes the trace
of the $2\times 2$ matrices.
 $\Tr_N$ is the trace over the $U(N)$ gauge indices and
$\Tr_\psi$ stands for the trace over the
 $SO(10,10)$ gamma matrices.
The symbol ${\rm Sym}$ in (\ref{CSterm}) means that we expand
$\wh C$ in the power series of $\Phi^\mu$ and
then symmetrize them with
$T^2$, $[T,\Phi^\mu]$, $[\Phi^\mu,\Phi^\nu]$.
We can show that the CS-term  (\ref{CSterm})
is invariant under the gauge transformation of the RR-fields
$C\rightarrow C+d\Lambda$,
generalizing the proof in \cite{OkOo} for the CS-term of BPS D-branes
including Myers' terms \cite{My}.
See Appendix \ref{App1} for the detail.

Taking the formal limit $N\rightarrow\infty$, we obtain the
Chern-Simons term for the IIA K-matrix theory. The trace $\Tr_N$
is replaced by the trace $\Tr_\cH$ over the Hilbert space $\cH$.

Let us consider a simple situation such that
 $C_{\mu_1\dots\mu_n}=\mbox{const.}$ ($n$ : odd)
is the only non-zero
RR-field, and $[\Phi^\mu,\Phi^\nu]=0$ for ${}^\forall \mu,\nu$.
Then, using the formula
\begin{eqnarray}
e^{A+B}&=&e^A+\int^1_0dt\, e^{(1-t)A}Be^{tA}\nonumber\\
&&+\int_0^1dt_1\int_0^{t_1}dt_2\,
e^{(1-t_1)A}Be^{(t_1-t_2)A}Be^{t_2A}+\cdots,
\end{eqnarray}
we can rewrite (\ref{CSterm}) as
\beqa
S_{\rm CS}
&=& C_{\mu_1\dots\mu_n}
\int_{0\leq t_n\leq\cdots\leq t_1\leq 1} dt_1 \cdots d t_n
\times\nonumber\\
&&~~~\Tr_{\cH}\left(
 e^{-(1-t_1) T^2} [T,\Phi^{\mu_1}] e^{-(t_1-t_2) T^2}
\cdots [T,\Phi^{\mu_n}] e^{-t_n T^2}
\right).
\label{CSterm2}
\eeqa
H\"older's inequality on the integrand in
(\ref{CSterm2}) implies that
\begin{eqnarray}
|S_{CS}|\leq \frac{1}{n!} |C_{\mu_1\dots\mu_n}|
\Tr_{\cH}\left(e^{-T^2}\right) \prod_{k=1}^n\norm{[T,\Phi^{\mu_k}]}.
\end{eqnarray}
Thus
the CS-term is
finite for the operators satisfying (\ref{finiteness}).
In general, the CS-term of a D-brane is estimated as
$|S_{CS}|\sim |C_{\mu_1\dots\mu_n}|V$, where $V$ is
the volume of the world-volume. Actually, as we will see
soon,  $\Tr_{\cH}\left(e^{-T^2}\right)$ is proportional to
the volume, and the first condition in (\ref{finiteness}),
which is called the $\theta$-summability condition,
is related to the compactness of the world-volume.
Thus, it could be relaxed if we allow the
infinite volume configurations.

Let us explain these facts in an explicit example, that is
the D-brane solution given in \cite{Te}.
We consider the Chan-Paton Hilbert space
 $\cH=L^2(\R^{2m+1})\otimes S$, where $S$ is $2^m$ dimensional
vector space of $SO(2m+1)$ spinors.
The D$(2m)$-brane configuration
is given by
\beqa
T &=& u\,D=u\,\sum_{\alpha=0}^{2m}
 \wh{p}_\alpha \otimes \gamma^\alpha,
\label{Dsol1}\\
\Phi^{\alpha} &=& \wh{x}^{\alpha} \otimes {\rm 1}
\;\; (\alpha=0, \cdots, 2m), \;\;\;\;
\Phi^i=0\;\; (i=2m+1, \cdots, 9),
\label{Dsol2}
\eeqa
where $\wh x^\alpha$
is defined by multiplication of $x^\alpha$
and $\wh p_\alpha=-i\partial/\partial x^\alpha$ is a differential
operator, both acting on $L^2(\R^{2m+1})$.
Note that the tachyon field $T$ is a Dirac operator $D$
up to normalization.
It can be shown that this becomes an exact BPS D$(2m)$-brane
 solution if we take $u\rightarrow \infty$, although this configuration
with finite $u$ still represents a D$(2m)$-brane.
Inserting this configuration into the Chern-Simons term (\ref{CSterm2})
with $n=2m+1$,
we obtain
\begin{eqnarray}
S_{\rm CS}=u^{2m+1}C_{01 \cdots 2m}\Tr_{\cH}\left(e^{-u^2 D^2}\right).
\end{eqnarray}
We can evaluated $\Tr_{\cH}\left(e^{-u^2 D^2}\right)$ as
\begin{eqnarray}
\Tr_{\cH}\left(e^{-u^2 D^2}\right)&=&2^m \int d^{2m+1}k\,
\langle k| e^{-u^2 k^2}|k\rangle\\
&=&
 \frac{\mu_{2m}}{u^{2m+1}}
\int d^{2m+1}x ,
\label{trace}
\end{eqnarray}
where $\mu_{2m}=1/(2^{m+1}\sqrt{\pi}^{2m+1})$ is a numerical constant.

Therefore the trace of the operator $e^{-u^2D^2}$ is proportional
to the volume factor $\int dx$ and diverges for the
infinite volume case.
In this paper, we will tacitly compactify the space-time
and consider $\theta$-summable family of tachyon operators
in order to avoid this divergence.

It is worthwhile to mention that the $S_{\rm CS}$ does not depend on
the parameter $u$ as expected, and the coefficient $\mu_{2m}$
gives the correct value of the coupling between the D$(2m)$-brane
and the RR $(2m+1)$-form field \cite{Te}.

\section{Spectral triples and D-branes}
\label{SpecTriple}

In section \ref{K-matrix}, we explained the possible configurations
of the K-matrix theory in purely analytic language.
In this section, we claim that these configurations correspond
 to the configurations of various D-branes embedded in the space-time,
and explain how to extract geometric information out of
the operators $\Phi^\mu$ and $T$ introduced in section \ref{IIA}.
In particular, as we will see in this section,
each configuration in the K-matrix theory
defines a spectral triple,
which is the analytic analog of the Riemannian manifold,
and we can use the techniques
developed in noncommutative geometry.

\subsection{Topology and algebra of D-branes}

Let $\wh\cA$ be the algebra generated by the operators
 $\Phi^\mu$ for a configuration in the IIA K-matrix theory.
Note that $\wh\cA$ is an involutive algebra of operators acting on a Hilbert
space $\cH$, since it is equipped with a star-operation
(Hermitian conjugation).
When $\wh\cA$ is a subalgebra of the bounded operator
algebra $\B(\cH)$ for the Hilbert space $\cH$,
$\wh\cA$ can be thought of as a $C^*$-algebra by taking the
completion.
\footnote{
A $C^*$-algebra is a norm closed self-adjoint subalgebra of
the bounded operator algebra $\B(\cH)$ for some Hilbert space $\cH$.}

Let us first consider the case that $\wh\cA$ is a commutative
$C^*$-algebra.
 The Gel'fand-Naimark theorem
states that every {\it commutative}
 $C^*$-algebra is of the form
$C_0(M)$, i.e. the space of continuous complex functions
on some locally compact Hausdorff space $M$, vanishing at infinity.
(If $M$ is compact, $C_0(M)$ is equal to $C(M)$ which is
 the space of continuous complex
functions on $M$. )
Note that the norm of an element of $C(M)$
is defined by its supremum value, and hence
there are unbounded elements in $C(M)$ if $M$ is not compact.

Moreover, it is known that
the category of commutative $C^*$-algebras
is in one-to-one correspondence with
the category of topological spaces
(locally compact Hausdorff spaces).
A point $x\in M$ on the space $M$
corresponds to a character of $\wh\cA$,
which is a *-homomorphism $\phi_x: \wh\cA=C_0(M)\rightarrow \C$.
Suppose that $\wh\cA$ is generated by mutually commuting operators
  $\Phi^\mu$ ($\mu=0,1,\dots,9$),
the character $\phi_x$ of $\wh\cA$ is given by picking up
one of the elements $x$ from the joint spectrum of
$(\Phi^0,\Phi^1,\dots,\Phi^9)$.
This agrees with the standard interpretation that
the eigenvalues of the matrix $\Phi^\mu$ represents
the position of the non-BPS D-instantons along the
space-time coordinate $x^\mu$.
When the spectrum of $(\Phi^0,\Phi^1,\dots,\Phi^9)$
agrees with the manifold $M$ embedded in $\R^{10}$,
we can say that $M$ is paved with non-BPS D-instantons.
Therefore,
 the topological space $M$ is interpreted as the
world-volume of higher dimensional objects made from
infinite number of non-BPS D-instantons.
Actually, as it will become clear in the next section,
$M$ is identified with the world-volume of higher dimensional
D-branes.

One major problem for this interpretation is that
$\Phi^\mu$ are not necessarily bounded operators and
the algebra generated by $\Phi^\mu$ may not be a
$C^*$-algebra. For example,
 if we take $\cH=L^2(\R^n)$
and $\Phi^\mu=\wh x^\mu$ ($\mu=1,\dots,n$),
which is the multiplication
operator $\Phi^\mu: f(x)\in \cH
\rightarrow x^\mu f(x)$,
the spectrum of $\Phi^\mu$ is not bounded
and hence $\Phi^\mu$ is an unbounded operator.
This happens when we consider the D-branes
with non-compact world-volume.
One way to avoid this problem is to compactify
the manifold $\R^n$ to $S^n$.
We can achieve this by replacing $\cH=L^2(\R^n)$ with
$\cH=L^2(S^n)$ and let
$\wh\cA$ be the algebra generated by
$\Phi^\mu=\wh x^\mu$ ($\mu=0,\dots,n$) with
a relation $\sum_{\mu=0}^n(\Phi^\mu)^2=R^2$,
that is $\wh\cA=C(S^n)$.
Another way is to restrict ourselves to the
subalgebra whose elements are of the form $f(\Phi^\mu)$
with some cut off function $f\in C_0(\R^n)$, and set
$\wh\cA=C_0(\R^n)$.
With these modification, we
assume that $\wh\cA$
consists of bounded operators and makes a $C^*$-algebra.

Another problem, which is a common issue in matrix theories,
is that the interpretation that the spectrum of $\Phi^\mu$
represent the coordinate of $x^\mu$ axis seems to be possible only when
the manifold has a global coordinate system
$(x^0,x^1,\dots,x^9)$ , (or its quotient such as
torus, orbifolds and so on).
It is not clear how to describe the theory when
the background manifold is topologically non-trivial.
An ad hoc resolution for this problem
is obtained by formally embedding the manifold
to a higher dimensional Euclidean space $\R^{N}$ ($N\ge 10$)
and introducing $\Phi^\mu$ ($\mu=0,\dots,N-1$)
 as the scalar field
corresponding to the fluctuation of $x^\mu$ direction,
which are subjected to some constraints representing the embedded
manifold.
More sophisticated description of the K-matrix theory
in general background
will be discussed elsewhere \cite{AsSuTe}.
(See also \cite{Do2,Do3,DoOoSh}.)

\subsection{Spectral triples}

In the last subsection, we saw that the $C^*$-algebra
$\wh\cA$ corresponds to a topological space $M$, which
is interpreted as the world-volume of D-branes.
Then, what is the geometric interpretation
of the tachyon operator $T$?
In this subsection, we claim that
the triple $(\cH,\wh\cA,T)$ can be interpreted as
the spectral triple, which is the basic ingredient for
noncommutative generalization of Riemannian geometry
\cite{Connes}.
In fact, the operator
$T$ gives the unit length scale of
the manifold $M$ and the infinitesimal line element $ds$ in
Riemannian geometry is identified with the
operator $1/|T|$.
Another significance of the operator $T$ is
that its homotopy class represents the K-homology
class of the manifold, which will be discussed
in section \ref{K-homo}.

Let us consider the triple $(\cH,\wh\cA,T)$, where
$\wh\cA$ is a $C^*$-algebra
generated by $\Phi^\mu$ acting on the
Chan-Paton Hilbert space $\cH$, and
$T$ is the (unbounded) tachyon operator, which
is a self-adjoint operator on $\cH$.
We assume here that $\wh\cA$ is unital,
i.e. $\wh\cA\ni\mbox{id}_\cH$, for simplicity.
For the commutative case, this means that we consider
the $C^*$-algebra $\wh\cA=C_0(M)=C(M)$  with compact
space $M$. Note that if
the topological space $M$ is non-compact,
$C_0(M)$ is not unital, since
the constant function with value 1
is not an element of $C_0(M)$.

Let us consider here the following conditions.
\begin{eqnarray}
(T-\lambda)^{-1}\in\K(\cH)~~
 \mbox{for}~{}^\forall \lambda\in\!\!\!\!\!/\,\R,
~~~[\wh a,T]\in\B(\cH)
~~ \mbox{for}~{}^\forall \wh a\in\wh\cA.
\label{spec}
\end{eqnarray}
The triple $(\cH,\wh\cA,T)$ satisfying these conditions
is called a spectral triple \cite{Connes}. (See also
\cite{Va,La}.)
The former condition in (\ref{spec}) means
that $T$ has a real discrete spectrum made of
eigenvalues $\{\lambda_n\in\R\}$
with finite multiplicity such that
$|\lambda_n|\rightarrow\infty$ as $n\rightarrow\infty$.
This is what we expect from the finiteness of the tachyon
potential, as we explained in section \ref{IIA}.
The latter one in  (\ref{spec}) is
nothing but the third condition in (\ref{finiteness}),
which is required from the finiteness of the tachyon kinetic
term.
Therefore the triple $(\cH,\wh\cA,T)$ defined by
a configuration of the K-matrix theory makes a spectral triple.

In general, we require a regularity hypothesis on
spectral triples $(\cH,\wh\cA,T)$ by replacing
$\wh\cA$ with a dense involutive
subalgebra of the $C^*$-algebra $\wh\cA$
\footnote{This subalgebra is
 generated by elements
$\wh a\in\wh\cA$ such that $\wh a$ and $[T,\wh a]$
are in the domain of $\delta^k$, where $\delta(f)=[|T|,f]$
is the derivation of the operator $f\in\B(\cH)$.}.
For example, we take $\wh\cA=C^\infty(M)$ instead of $C(M)$
for the commutative cases.

A basic example of the spectral triple, which is
called the canonical triple, is given by
$(\cH,\wh\cA,T)=(L^2(M,S),C^\infty(M),D)$.
Here $M$ is a closed Riemannian spin manifold
\footnote{One can also construct a canonical triple
over a spin${}^c$ manifold, by adding a $U(1)$ gauge
connection.},
$L^2(M,S)$ is
the Hilbert space of square integrable sections
of the spinor bundle on $M$, and
 $D$ is the Dirac operator associated with the Levi-Civita
connection of the metric.
This is essentially nothing but the D-brane configurations
considered in section \ref{CSandD} for the case $M=\R^{2m+1}$,
though we assumed $M$ to be compact and could be curved
here.
Here, we take $u=1$ for the normalization of
the tachyon in (\ref{Dsol1}),
so that $\norm{[\Phi^\mu,T]}=1$.
As explained in the previous subsection, a point $x$
of $M$ is given by the character $\phi_x$ of $\wh\cA=C(M)$,
and $x^\mu = \phi_x (\Phi^\mu)$ for the flat $M=\R^n$ case.
Therefore the distance between two points $x_1,x_2\in \R^n$
 is given by $|\vec{x}_1-\vec{x}_2|=
|\phi_{x_1}(\vec{\Phi})-\phi_{x_2}(\vec{\Phi})|$,
where $\vec{x}_1 =(x_1^1,\dots,x_1^n)$ etc.
Then, it is not hard to imagine that this can be generalized as
\begin{eqnarray}
d(\phi_1,\phi_2)=\sup_{a\in\wh\cA}\left\{~|\phi_1(a)-\phi_2(a)|~
\Big|~~\norm{[T,a]}\le1~\right\},
\label{distance}
\end{eqnarray}
where $\phi_i$ ($i=1,2$)
are linear functions $\phi_i:\wh\cA\rightarrow\C$
such that $\phi_i(a^*a)\ge 0$ for ${}^\forall a\in\wh\cA$
and normalized as $\phi_i(1)=1$. Such functions as $\phi_i$
are called states and the distance $d(\phi_1,\phi_2)$
between two states in an arbitrary spectral triple
is defined by this formula.
It is known that this agrees with the geodesic distance
between two points for the canonical triples,
when we take $\phi_i$ as characters that represent
the two points. (See for example \cite{Connes,La}).

In this way, the operator $T$ carries information about
the metric on the world-volume of the D-brane.
More explicitly,
for the canonical triple, the asymptotic expansion
of the heat kernel of $T^2$ at small $t$ is known as
\cite{BrGi}
\begin{eqnarray}
\Tr_{\cH}\left(e^{-t\,T^2}\right)\sim
\frac{2^{[n/2]}}{(4\pi t)^{n/2}}\int_M d^nx\sqrt{g}
 \left(1+\frac{t}{12} R+O(t^2)\right),
\label{exp}
\end{eqnarray}
from which we can measure
the volume of the world-volume, integral of
the mean curvature and so on.
Note that
the first term in the expansion is used in (\ref{trace})
to derive the CS-term of the D-brane.

Note that
the metric defined in (\ref{distance}) is, in general,
different from the usual metric of the D-brane
induced from the background metric via the embedding, since
the metric defined by the tachyon operator depends on
the scale of the tachyon condensation. Namely, the unit length
is defined by the scale of the tachyon condensation.
Anyway, the action of the higher dimensional D-brane
represented by the spectral triple $(\cH,\wh\cA,T)$,
should be written covariantly using the world-volume
metric defined by the tachyon,
which is analogous to the covariant
Polyakov action in string theory.

Let us next explain the fact that information of
the dimension of the D-brane world-volume is also
hidden in the spectrum of $T$.
As we can see in the expansion (\ref{exp}),
the dimension $n$ of the space $M$ can be read
from the power of $t$ in the right hand side.
More generally,
the notion of dimension of a spectral triple is
replaced by dimension spectrum which is a subset
$\Sigma\subset\C$ of the singularities of the analytic
function
\footnote{More precisely, the dimension spectrum is the singularities of
$\zeta_{b}(z)=\Tr_{\cH}(b|T|^{-z})$
where $b$ is an element of the algebra generated by
$\delta^k(\wh a)$, $\delta^k([T,\wh a])$ with
$\wh a\in\wh\cA$.}
\cite{CoMo,Connes}
\begin{eqnarray}
\zeta_T(z)=\Tr_{\cH}(|T|^{-z}).
\end{eqnarray}
It is easy to show that it gives the dimension $n$
for the expansion (\ref{exp}), as expected,
using the Mellin transform
\begin{eqnarray}
\Tr_{\cH}(|T|^{-z})=\frac{1}{\Gamma(z/2)}\int_0^\infty
t^{z/2-1}\Tr_{\cH}(e^{-tT^2})dt.
\end{eqnarray}
Interestingly,
there are some examples that the dimension defined
above does not take an integer value \cite{Connes,GuIs}.
It would be interesting if
a fractal D-brane is realized in the K-matrix theory.

Now we discuss the diffeomorphism of the spectral triples.
In \cite{Connes}, the diffeomorphism of the geometry
represented by the spectral triple and the local gauge transformation
on it are discussed.
Then in the K-matrix theory,
these may be realized in the spectral triples which represent
the world-volume of the D-branes.
Actually, a subset of the unitary operators in $\B(\cH)$ can be
interpreted as $\{ \mbox{local gauge transf.} \times
\mbox{diffeo.} \}$ in the explicit examples of D-brane configurations
with commutative world-volumes.
Let us explain this below.

A configuration of curved $N$ D$(2m)$-branes with the $U(N)$ gauge field
is given by
\beqa
\label{curvedD}
\Phi^{\mu} &=& f^{\mu} ( \hat{x}^{i} )  \;\;\; (i=0, \ldots, 2m) \CR
T & = &\half\left\{ \gamma^a e^{i}_{\, a} (\hat{x}) ,
\left( \hat{p}_i +w_i^{ab}(\hat{x})\gamma_{ab} +A_i(\hat{x})
\right)
\right\},
\eeqa
where $[\hat{x}^i, \hat{x}^j ]=0 $,
$[\hat{x}^i, \hat{p}_j ]=i\delta^i_j $
 and $\{ \gamma^a, \gamma^b \}= 2 \eta^{ab}$.
Here we think that $A_i(\hat{x})$ is the $U(N)$ gauge field,
$f^\mu(\hat{x})$ is the embedding function,
$e^i_{\, a} (\hat{x})$ are vielbein and
$w_i^{ab}(\hat{x})$ is the spin connection constructed from them.
Remember that the configurations of the theory
should be physically identified with
each other by the unitary transformations,
since they are the gauge transformation of the
underlying non-BPS D-instantons.
However it is difficult to obtain
the usual geometric picture
for the transformed configurations,
if $\Phi^{\mu}$ in the transformed configuration
depend on $\hat{p}^i$, $\gamma^i$ or $N \times N$ matrices.
Then in order to interpret the transformations
in terms of the usual geometric picture,
we will consider the unitary operators $u=e^{i H}$, where
\beq
H= \lambda(\hat{x},\gamma^i)+\half \{ \hat{p_i},  \epsilon^i (\hat{x}) \},
\label{transf}
\eeq
$\epsilon^i$ is an identity as an $N \times N$ matrix,
and $\lambda$ is an $N \times N$ matrix valued function.

Since these operators form a subgroup and
the action of the K-matrix theory is invariant under these transformations,
we can in principle obtain an action for the D$(2m)$-branes, which is
invariant under these transformations,
by evaluating the action of the K-matrix theory including
all the fluctuations around the configuration (\ref{curvedD}).
However, the fields on the
$N$ D$(2m)$-branes in (\ref{curvedD})
do not include all the fluctuations.
Indeed, there are the $n$-form fields associated to
the $\gamma^{i_1} \cdots \gamma^{i_n}$, for example,
and the transformation using $\lambda(\hat x,\gamma^i)$
 includes the $n$-form gauge transformation.

Here we consider the fields in (\ref{curvedD}) only and
the transformations generated by (\ref{transf}) with
$\lambda=\lambda(\hat{x})$ because these transformations
form a subgroup and act consistently on these fields.
Then the transformation using $ u= \exp (i \lambda(\hat{x}))$
corresponds to the local gauge transformation since
$u \Phi^{\mu} u^{-1}
=\Phi^{\mu}$ and $u T u^{-1} = \gamma^a e^{i}_{\, a} (\hat{x})
\left( \hat{p}_i +w_i(\hat{x}) +A^{\lambda}_i(\hat{x}) \right) $,
where $A^{\lambda}_i=u A_i u^{-1}-i u\pa_i u^{-1}$.
The transformation using
$ u_d= \exp (i \half \{ \hat{p}_i , \epsilon^i(\hat{x}) \})$
corresponds to the diffeomorphism of the world-volume
of the D$(2m)$-branes.
For simplicity we assume that $\epsilon^i$ is infinitesimal.
Then we can verify that
$u_d \Phi^{\mu} u_d^{-1} = f^{\mu} (\hat{y}(\hat{x})) $ and
$u_d T u_d^{-1} =\half\left\{ \gamma^a {e'}^{i}_{\, a} (\hat{x}),
\left( \hat{p}_i +{w'}_i^{ab} (\hat{x} ) \gamma_{ab}
+A'_i(\hat{x}) \right) \right\}$,
where we set $\hat{y}^i=\hat{x}^i+\epsilon^i(\hat{x})$ and
\begin{eqnarray}
e'^i_{\, a}=
\frac{\pa \hat{x}^i}{\pa \hat{y}^j} e^j_{\, a}(\hat{y}(\hat{x})) ,~~
\hat{p}_i=-i \frac{\pa}{\pa \hat{x}^i},~~
{w'}_i^{ab}= \frac{\pa\hat y^j}{\pa\hat x^i} w_j^{ab}(\hat{y}(\hat{x})),
~~A'_i= \frac{\pa\hat y^j}{\pa\hat x^i} A_j(\hat{y}(\hat{x})).
\end{eqnarray}

Therefore the world-volume theory of the D$(2m)$-branes,
which is the K-matrix action evaluated by (\ref{curvedD}),
has the invariance under the diffeomorphism and local gauge
transformation.
Indeed, the large metric expansion of the K-matrix action
is a Polyakov type action
\beq
S \sim \int dx^{2m+1} \sqrt{g} \left(
1+ 2 \log 2 G_{\mu \nu} \pa_i f^\mu \pa_j f^\nu g^{ij} +\cdots \right),
\eeq
where $g^{ij}=e^i_{\, a} \eta^{ab} e^j_{\,b}$
is the world-volume metric and $G_{\mu \nu}=\eta_{\mu \nu}$ is
the background metric.

In the case of noncommutative D-brane configuration,
we may also be able to
give an interpretation to the unitary transformation
as $\{ \mbox{local gauge transf.} \times
\mbox{diffeo.} \}$ as above.

\subsection{Embedding of D-branes}
\label{embed}

The spectral triples considered in the previous subsection
represent the geometry of the world-volume of D-branes,
and we did not specify the space-time manifold in which
the D-branes are embedded.
In this subsection, we fix a space-time manifold $X$,
and explain how to describe D-branes embedded in $X$
in the algebraic description.
We will use this set up for the classification of D-branes
in the next section.

Let us recall the correspondence between
the category of topological spaces
(locally compact Hausdorff spaces) and
the category of algebras ($C^*$-algebra).
(See for example \cite{Va}.)
\begin{center}
\begin{tabular}{ccccc}
\hline\hline
{\bf Topology}&$\lra$&{\bf Algebra}&:& commutative case\\
\hline
topological space $X$&$\lra$& $C^*$-algebra $\cA$&:& $C_0(X)$\\
compact&$\lra$& unital&:& $C_0(X)=C(X)\ni 1$\\
proper map $\varphi: M\rightarrow X$
&$\lra$& *-homomorphism &:& $\varphi^*:C_0(X)\rightarrow C_0(M)$
\\
homeomorphism&$\lra$& automorphism&:&\\
open subset $U\subset X$&$\lra$& ideal &:& $J_{X-U}$\\
closed subset $V\subset X$&$\lra$& quotient algebra &:&$C_0(X)/J_V$
\\
\hline
\end{tabular}
\end{center}
Here the ideal $J_V$ associated with a closed subset $V\subset X$ is
defined as $J_V=\{\,f\in C_0(X)\,|\,\,f|_V=0\,\}$.
The proper map $\varphi$ is a continuous map such that
the inverse image $\varphi^{-1}(K)$
of any compact subset $K$ in $X$ is compact.
In other words, roughly speaking,
$\varphi$ maps infinity to infinity.
Note that $\varphi^* f=f\circ\varphi$ for
a function $f\in C_0(X)$ may not
vanish at infinity, if $\varphi$ is not
a proper map.

Let us consider a D-brane world-volume $M$
embedded in the space-time manifold $X$.
Let $\cA=C_0(X)$ and $\wh\cA=C_0(M)$ be the algebra
corresponding to $X$ and $M$ respectively.
Since $M$ is a closed subset of $X$ and
does not have a boundary except infinity,
the inclusion map $i:M\rightarrow X$ should be
a proper map. Therefore the inclusion map induces
a *-homomorphism $i^* : \cA\rightarrow \wh\cA
\simeq \cA/J_M$.
The ideal $J_M$ is the kernel of $i^*$ by definition.
The generalization to the noncommutative cases is straightforward.
Let $\cA$ be the $C^*$-algebra that corresponds to the
space-time manifold, which could be noncommutative.
In order to obtain an algebra $\wh\cA$ corresponding
to the world-volume of the D-brane embedded in the
space-time, we choose a *-homomorphism
$\phi:\cA\rightarrow\B(\cH)$ and set
$\wh\cA=\mbox{Image}\,\phi\cong\cA/\ker\phi$.
For example, suppose that the space-time algebra is
$\cA=C(X)$, where $X$ is a compact subset of $\R^N$,
and a *-homomorphism
$\phi:\cA\rightarrow\B(\cH)$ is given,
 $\wh\cA$ is defined as
the algebra generated by $\Phi^\mu=\phi(x^\mu)$,
where $(x^0,\dots,x^{N-1})$ is the coordinate of $\R^N$.

As emphasized in \cite{IIBreview},
we do not apriori have the notion of space-time manifold
in matrix theory.
As an example, consider the IIA K-matrix theory
formulated in the flat $\R^{10}$ background.
The eigenvalues of the scalar field $\Phi^\mu$
represent the positions of the non-BPS D-instantons
in $x^\mu$ direction of the $\R^{10}$.
However, there are configurations that $\Phi^\mu$
are mutually noncommutative and they cannot be simultaneously
diagonalized. In such cases, we cannot say
that the non-BPS D-instantons live in the $\R^{10}$ space-time.

But, it is still interesting to consider the possible configurations
of D-branes embedded in a fixed space-time manifold
using the framework of the matrix theory.
For instance, if we are only interested in the commutative D-branes
embedded in $\R^{10}$, it is reasonable to fix the space-time
algebra as $\cA=C_0(\R^{10})$
and consider only the D-branes represented by
the algebra $\wh\cA =\mbox{Image}\,\phi$ for some *-homomorphism
$\phi:\cA\rightarrow \B(\cH)$. Note that, in this case,
 we can never obtain D-brane configurations with noncommutative
world-volume algebra $\wh\cA$ as the image of $\phi$,
since $\cA=C_0(\R^{10})$ is
commutative and the map $\phi$ is homomorphism.

In the next section, we will first fix a $C^*$-algebra $\cA$,
which we call the space-time algebra, and
classify the stable D-brane configurations which are embedded
in the space-time represented by $\cA$.
Then, D-branes embedded in the space-time are
represented by the spectral triples $(\cH,\wh\cA,T)$,
where $\wh\cA$ is given by $\wh\cA=\mbox{Image}\,\phi$,
using a *-homomorphism $\phi:\cA\rightarrow \B(\cH)$.
In other words, such D-branes are obtained by
a triple $(\cH,\phi,T)$, which is called an (unbounded)
Fredholm module.
Therefore,
the classification of D-brane configurations
are obtained by classifying the Fredholm modules,
which we will demonstrate in the next section.

\section{D-branes and K-homology}
\label{K-homo}

\subsection{Classification of D-brane configurations}

The D-brane charge is classified by K-theory group $K^1(X)$ ($K^0(X)$)
in type IIA (IIB) string theory \cite{Wi,Ho}. So,
at first sight, one might think that the D-brane configurations
in the K-matrix theory
should be naturally
classified by the algebraic K-theory $K_1(\cA)$ ($K_0(\cA)$),
since they are isomorphic to the topological
K-theory $K^i(X)=K_i(C(X))$ when $\cA=C(X)$.
However, it turns out that this is {\it not} the correct answer.
Let us explain this fact shortly.
\footnote{A similar argument is given in \cite{Pe}.}

The charge of D-branes is usually defined by the behavior of
RR-fields. Therefore, it should be classified by
cohomology theory. Actually, K-theory is a kind of refined cohomology theory.
In particular, it behaves as a contravariant functor
from the category of topological spaces to the category of
 Abelian groups.
This means that
a diffeomorphism $\phi:X\rightarrow X'$ induces a pull-back map
\begin{eqnarray}
\phi^*:K^i(X')\rightarrow K^i(X).
\end{eqnarray}

On the other hand, when we construct D-branes
in the K-matrix theory with $\cA=C(X)$,
the D-brane solutions
represent cycles of $X$ which correspond
to the world-volume of the D-branes. Hence, they should be
classified by homology theory, which is Poincare dual to
the cohomology theory and
transforms covariantly under the diffeomorphism $\phi$.

There is a group called K-homology $K_i(X)=K^i(C(X))$, which
is dual to the K-theory group $K^i(X)=K_i(C(X))$ in the sense that
it has a natural pairing with the K-theory group,
\begin{eqnarray}
K^i(X)\times K_i(X)\rightarrow \Z.
\end{eqnarray}
Accordingly, the K-homology $K_i(X)$ is a homological object, which
is preferable to our purpose.
In the next subsection, we claim that
the K-homology $K^1(\cA)$ ($K^0(\cA)$) is the group which classifies
the D-brane configurations in the type IIA (IIB) K-matrix theory.
(See also \cite{HaMo,Pe,Sz}.)

\subsection{Analytic K-homology}
\label{analyK}

There are several ways to define the K-homology.
(See, for example, \cite{BaDo,Bl,MaSi}.)
One of them is the definition of the K-homology using Fredholm
operators. We can also define it in terms of manifolds and
vector bundles when the algebra $\cA$ is commutative.
As we will see in this subsection,
the former have a direct physical interpretation
in the K-matrix theory, since our formulation is based on
the operator algebra.
The latter topological approach is useful to
relate the elements of K-homology to world-volume
configurations of D-branes,
and will be discussed in the next subsection.

First we fix the space-time $C^*$-algebra $\cA$.
We assume $\cA$ to be unital for simplicity.
As explained in section \ref{embed},
the D-branes embedded in the fixed space-time algebra $\cA$ are
obtained by the Fredholm modules.
A Fredholm module over an algebra $\cA$
is a triple ($\cH,\phi,F$), where
\begin{itemize}
\item $\cH$ is a separable Hilbert space,
\item $\phi: \cA \rightarrow \B(\cH)$ is a *-homomorphism,
\item $F$ is a self adjoint operator in $\B(\cH)$,
 which satisfies
\begin{eqnarray}
 F^2-1 \in \K(\cH),~~~
 [F,\phi(a)] \in {\K(\cH)}~~\mbox{for}~{}^\forall a \in \cA.
 \label{cptness2}
\end{eqnarray}
\end{itemize}
As explained in section \ref{embed},
a Fredholm module $(\cH,\phi,F)$ describes a
configurations of the IIA K-matrix theory.
$\cH$ is identified as a space of Chan-Paton indices
of the non-BPS D-instantons,
the *-homomorphism $\phi$
specifies the world-volume $\wh\cA=\mbox{Image}~\phi$ of the D-branes
embedded in the space-time algebra $\cA$,
and the operator $F$ is
the normalized tachyon field $T_b$.
The condition  (\ref{cptness2}) is nothing but the condition
(\ref{cptness}), which is required from the finiteness of the action.

We also define a degenerate Fredholm module
which is a Fredholm module
satisfying $F^2-1= [F,\phi(a)] =0$.
This corresponds to virtual non-BPS D-instantons
that would be annihilated by the tachyon condensation.
The sum of two Fredholm modules
($\cH_i,\phi_i,F_i$) ($i=0,1$)
 are defined by the direct sum
($\cH_0\oplus\cH_1,\phi_0\oplus\phi_1,F_0\oplus F_1$).

Two Fredholm modules ($\cH_i,\phi_i,F_i$) ($i=0,1$)
are said to be unitary equivalent when there is a unitary operator in
$\B(\cH_0,\cH_1)$ intertwining $\phi_i$ and $F_i$.
They are operator homotopic if $\cH_0=\cH_1$, $\phi_0=\phi_1$ and
there is a norm continuous path between $F_0$ and $F_1$.
We define an equivalence relation $\sim$ on Fredholm modules
generated by unitary equivalence,
addition of degenerate elements and operator homotopy
of ($\cH,\phi,F$).
Then K-homology $K^1(\cA)$ is defined as the set of equivalence classes
of the Fredholm modules under the equivalence relation $\sim$.

These equivalence relations have nice physical interpretations.
The unitary equivalence is nothing but the
gauge equivalence,
addition of the degenerate elements means the
addition of non-BPS D-instantons that would be
annihilated by the tachyon condensation, and the operator homotopy
is just a continuous deformation of the tachyon configuration.
One could also consider the continuous deformation of
$\cH$ and $\phi$, though it is known that the equivalence class
is unchanged \cite{Bl}. Therefore the equivalence
relations considered above is physically
enough for the classification of the configurations.

The K-homology which classifies the D-brane configurations
in the IIB K-matrix theory is $K^0(\cA)$.
It can be defined in a similar way.
In this case, a Fredholm module over an algebra $\cA$
is a 5-tuple $(\cH^{(0)},\cH^{(1)},\phi_0,\phi_1,F)$, where
\begin{itemize}
\item $\cH^{(i)}$
 are separable Hilbert spaces ($i=0,1$),
\item $\phi_i: \cA \rightarrow \B(\cH^{(i)})$
 are  *-homomorphisms ($i=0,1$),
\item $F$ is an operator in $\B(\cH^{(0)},\cH^{(1)})$, which satisfies
\begin{eqnarray}
&&F^*F-1 \in \K(\cH^{(0)}),
~~FF^*-1 \in \K(\cH^{(1)}),
\label{Pot2}\\
&& F\phi_0(a)-\phi_1(a)F \in {\K(\cH^{(0)},\cH^{(1)})}
 ~~\mbox{for}~ {}^\forall a \in \cA.
\label{Tkin2}
\end{eqnarray}
\end{itemize}
This Fredholm module
$(\cH^{(0)},\cH^{(1)},\phi_0,\phi_1,F)$
 describes the configurations of the IIB K-matrix theory
 in an analogous way as above.
$\cH^{(0)}$ and $\cH^{(1)}$
corresponds to the Chan-Paton indices of D-instantons
and anti D-instantons, respectively.
The *-homomorphisms $\phi_0$ and $\phi_1$
is used to obtain the configurations
that D-instantons and anti D-instantons are
settled inside the space-time manifold in the same way as explained
in section \ref{embed} for non-BPS D-instantons.
$F$ is again the normalized tachyon field $T_b$.
Then, (\ref{Pot2}) and (\ref{Tkin2}) are the conditions
corresponding to
(\ref{Pot}) and (\ref{Tkin}), respectively.

The degenerate Fredholm module
is defined as the 5-tuple
 ($\cH^{(0)},\cH^{(1)},\phi_0,\phi_1,F$)
with $\cH^{(0)}=\cH^{(1)}$, $\phi_0=\phi_1$ and
$F=\mbox{id}_\cH$ is the identity operator of $\cH\equiv\cH^{(0)}=\cH^{(1)}$.
The K-homology $K^0(\cA)$ is defined as the set of equivalence classes
of the Fredholm module.
The equivalence relations of the Fredholm modules are
again generated by unitary equivalence, addition of degenerate
elements and the operator homotopy.

We can also rewrite these conditions in terms of
the $\Z_2$ graded Hilbert space $\wh\cH=\cH^{(0)}\oplus\cH^{(1)}$.
The conditions (\ref{Pot2}) and (\ref{Tkin2}) can be expressed as
\begin{eqnarray}
\wh F^2-1\in \K(\wh\cH),~~~[\wh\phi(a),\wh F]\in \K(\wh\cH),
~~~ \mbox{for}~{}^\forall a\in\cA,
\end{eqnarray}
where $\wh\phi(a)=\diag(\phi_0(a),\phi_1(a))$ and
$\wh F=\left({~~~F^*\atop F~~~} \right)$.

When $\cA=C(X)$, where $X$ is a compact manifold,
there is a surjective map from $K^0(C(X))=K_0(X)$ to $\Z$,
\begin{eqnarray}
\Index: K_0(X)\rightarrow \Z.
\label{index}
\end{eqnarray}
This map is defined by taking the index of the Fredholm operator $F$,
\begin{eqnarray}
\Index((\cH^{(0)},\cH^{(1)},\phi_0,\phi_1,F)) \equiv \Index F.
\end{eqnarray}
The index of a Fredholm operator is invariant under
operator homotopy and the map (\ref{index}) is well defined.
Recall that $F$ is the tachyon field of the IIB K-matrix theory
and gives a map from D-instanton
Chan-Paton Hilbert space $\cH^{(0)}$
to anti D-instanton Chan-Paton Hilbert space $\cH^{(1)}$.
The basis of $\mbox{Ker}\,F$ and $\mbox{Coker}\,F$ correspond to
the Chan-Paton indices for the D-instantons and anti D-instantons
 which are not annihilated by
the tachyon condensation, respectively. Therefore
the integer $\Index F = \mbox{Ker}\,F - \mbox{Coker}\,F$
 is interpreted as the total number
of D-instantons.
Note that
we can realize the configurations with any numbers of D-instantons
in the K-matrix theory.
This is one of the advantage of the K-matrix theory
in contrast to the other matrix theories.

Let us examine $\cA=C(S^n)$ case as a basic example.
K-homology groups for this algebra are known as
\begin{eqnarray}
K_0(S^n)=\left\{
\begin{array}{cc}
\Z\oplus\Z&(n:\mbox{even})\\
\Z&(n:\mbox{odd}),
\end{array}
\right.
~~~~K_1(S^n)=\left\{
\begin{array}{cc}
0&(n:\mbox{even})\\
\Z&(n:\mbox{odd}).
\end{array}
\right.
\end{eqnarray}
These results are consistent with what we expect from
the homology group of $S^n$.
$S^n$ has non-trivial homology for
$H_0(S^n)=\Z$ and $H_n(S^n)=\Z$, and hence the only
topologically non-trivial D-branes
wrapped on $S^n$ are expected to be
D-instantons and D$(n-1)$-branes. Here $n$ should be
odd (even) in type IIA (IIB) string theory  to obtain
a stable D-branes wrapped on $S^n$.
Since we always have the $\Z$ factor corresponding
to the D-instantons, it is convenient to consider
the reduced K-homology group, defined as the kernel
of the index map (\ref{index}), or equivalently
replacing $C(S^n)$ with $\cA=C_0(\R^n)$.
\begin{eqnarray}
K_0(\R^n)=\left\{
\begin{array}{cc}
\Z&(n:\mbox{even})\\
0&(n:\mbox{odd}),
\end{array}
\right.
~~~~K_1(\R^n)=\left\{
\begin{array}{cc}
0&(n:\mbox{even})\\
\Z&(n:\mbox{odd}).
\end{array}
\right.
\end{eqnarray}
Here $K_i(\R^n)$ denote $K^i(C_0(\R^n))$.
In this case,
since the D-instantons can be kicked off to infinity,
the $\Z$ factor corresponding to the D-instanton number
is dropped.
Therefore we could say that
the flat D$p$-brane is classified by
$K_1(\R^{p+1})$ ($K_0(\R^{p+1})$) in the IIA (IIB) K-matrix theory.

Note that in the K-theory classification of D-brane charges
\cite{Wi}, $K^i(\R^{9-p})$
\footnote{Here $K^i(\R^{n})$ denotes
the reduced K-theory group of $S^n$, $\widetilde K^i(S^n)$
\cite{Ka}.}
is the group which classifies the
charge of the flat D$p$-branes considered above.
The space $\R^{9-p}$ is the dual of $\R^{p+1}$
in the space-time $\R^{10}$.
This comes from the fact that K-theory of $X$ classifies
the D-brane charge defined by RR-fields on $X$,
 while K-homology of $X$ classifies
the world-volume of the D-brane embedded in the
space-time manifold $X$, and they are related by
Poincar\'e duality.

In general, for a $n$-dimensional
 compact manifold $X$,
 the K-theory groups and the K-homology groups are related by
\begin{eqnarray}
K_i(X)\simeq K^{n-i}(X),
\label{Kisom}
\end{eqnarray}
where the subscript $i$ and the
superscript $n-i$ are understood modulo 2.
This isomorphism is the K-theory lift of
the Poincar\'e duality
\begin{eqnarray}
H_i(X;\Z)\simeq H^{n-i}(X;\Z).
\end{eqnarray}
The isomorphism (\ref{Kisom}) is also
reasonable from the physical point of view.
For example, in type IIA string theory,
$K_1(X)$ classifies the D-brane constructed by
non-BPS D-instanton system and
$K^{n-1}(X)$ classifies the D-brane constructed by
non-BPS D$(n-1)$-brane system when $n$ is even,
or D$(n-1)$-brane - anti D$(n-1)$-brane system when $n$ is odd.
The spectrum of the D-branes should not depend on how
they are constructed, and hence $K_1(X)$ and $K^{n-1}(X)$
should be isomorphic. We will generalize this discussion
in section \ref{KK-theory}.

\subsection{Topological K-homology}
\label{topK}

When the algebra $\cA$ is commutative, we have a
topological definition of the K-homology, which is
isomorphic to the analytic one given in
the previous subsection.
(See \cite{BaDo,HaMo,MaSi}.)
In this case, we can assume $\cA=C_0(X)$ without
any loss of generality. Here $X$ can be
any locally compact Hausdorff space,
but, for simplicity, we assume that $X$ is a closed (i.e. compact
without boundaries)
smooth manifold in this subsection.

A K-cycle on $X$ is
defined to be a triple $(M,E,\varphi)$, where
$M$ is a compact Spin${}^c$ manifold without boundary,
$E$ is a complex vector bundle on $M$, and
$\varphi$ is a continuous map from $M$ to $X$.
Note that we don't require
the manifold $M$ to be connected, and the rank of $E$ may
be different on different connected components of $M$.
Therefore, the disjoint union $(M_0,E_0,\varphi_0)\cup
(M_1,E_1,\varphi_1)$ of two K-cycles $(M_i,E_i,\varphi_i)$ $(i=0,1)$
is again a K-cycle.

The (topological) K-homology
$K_*^{top}(X)=K^*_{top}(C(X))$
is the set of equivalence classes of the K-cycles.
The equivalence relations are generated by the following
(a)$sim$(c):
\begin{enumerate}
\item Bordism

$(M_0,E_0,\varphi_0)\sim(M_1,E_1,\varphi_1)$~~ if there exists
a triple $(W,E,\varphi)$,
such that $(\partial W, E|_{\partial W},\varphi|_{\partial W})$
is isomorphic to the disjoint union
 $(M_0,E_0,\varphi_0)\cup(-M_1,E_1,\varphi_1)$.
Here $W$ is a compact Spin${}^c$ manifold with boundary,
$E$ is a complex vector bundle on $W$,
$\varphi$ is a continuous map from $W$ to $X$,
and $-M_1$ denotes $M_1$ with the reversed Spin${}^c$ structure.

\item Direct sum

$(M,E_1\oplus E_2,\varphi)\sim(M,E_1,\varphi)\cup (M,E_2,\varphi)$

\item Vector bundle modification

$(M,E,\varphi)\sim(\wh M,\wh H\otimes\rho^*(E),\varphi\circ\rho)$,
~~where $\wh M$ is a sphere bundle on $M$ whose
fiber $S_p$ is an even dimensional sphere, $\rho$ is the
projection $\wh M\rightarrow M$ and $\wh H$ is a vector
bundle on $\wh M$, such that for each  $p\in M$
 the restriction of $\wh H$ to $S_p=\rho^{-1}(p)$
is the generator of $\widetilde K(S_p)=\Z$.
(See \cite{BaDo} \S10 for the explicit construction.)

\end{enumerate}

The sum of two elements in the K-homology is defined by
the disjoint union, and it can be shown that $K_*^{top}(X)$
is an Abelian group. $K_*^{top}(X)$ is a direct sum of
two subgroups $K_i^{top}(X)$ ($i=0,1$),
\begin{eqnarray}
K_*^{top}(X)= K_0^{top}(X)\oplus K_1^{top}(X),
\end{eqnarray}
where $K_0^{top}(X)$ ($K_1^{top}(X)$) consists of
the elements given by the K-cycles $(M,E,\varphi)$
with each component of $M$ even (odd) dimensional.

It is natural to interpret the K-cycle $(M,E,\varphi)$
as the world-volume of the D-brane as proposed in
\cite{HaMo}. $M$ is interpreted as the world-volume of the
D-brane with Chan-Paton bundle $E$ on it and $\varphi$
determines the embedding of the D-brane to the space-time $X$.
The requirement that $M$ is
equipped with a Spin${}^c$ structure is
consistent with the fact that D-branes cannot
wrap on a cycle without any Spin${}^c$ structures
\cite{FrWi,Wi3,Wi,BrSh}.

The topological K-homology $K_1^{top}(X)$ $(K_0^{top}(X))$
 defined above nicely
classifies the stable D-brane configurations
in type IIA (IIB) string theory.
The equivalence relation (a) is the deformations of
the world-volume of the D-brane together with the gauge bundle on it,
the relation (b) represents the process
of the gauge symmetry enhancement for coincident
D-branes.
The relation (c) is the descent relation of
the D-branes, namely it means that
we should identify
a spherical D-brane with a non-trivial gauge bundle on it
with a lower dimensional D-brane.
 Let us explain this fact in a little more detail.
The Spin${}^c$ manifold $\wh M$ and the
vector bundle $\wh H$ are constructed as follows.
Let $H$ be a Spin${}^c$ vector bundle on $M$ with $2n$ dimensional
fibers, and $B(H)$ be the unit ball bundle of $H$.
The boundary of $B(H)$ is a unit sphere bundle $S(H)$ on $M$,
whose fiber is $2n-1$ dimensional sphere.
$\wh M$ is defined by gluing
 two copies of $B(H)$, denoted by $B(H)_+$ and $B(H)_-$,
by the identity map of $S(H)$.
 Thus, $\wh M$ is a sphere
bundle on $M$ with $2n$ dimensional sphere
as its fiber, and is also
 a Spin${}^c$ manifold.
$B(H)_+$ and $B(H)_-$ are regarded as the world-volume of
a D$(2n+p)$-brane and an anti D$(2n+p)$-brane, respectively,
where $p=\dim M-1$. Gluing them together, $\wh M$
can be thought of as the world-volume of a spherical
 D$(2n+p)$-brane, wrapped on $2n$ dimensional sphere.
Since the fiber of the Spin${}^c$ vector bundle $H$
is even dimensional,
we can define two spinor bundles $H_\pm$,
labeled by chirality, on $M$ associated to $H$.
Let $S_\pm$ be the pull-backs of $H_\pm$ to $H$,
using the projection $H\rightarrow M$.
We associate $S_+$ and $S_-$ as the Chan-Paton vector
bundle on the D$(2n+p)$-brane and the anti D$(2n+p)$-brane,
respectively.
We restrict the base $H$ of $S_\pm$ to $B(H)_\pm$
and denote them by $S_\pm|_{B(H)_\pm}$.
The vector bundle $\wh H$ on $\wh M$ is constructed by gluing
$S_+|_{B(H)_+}$ and $S_-|_{B(H)_-}$
 by the transition function $g$ on
$S(H)$. The transition function $g$ is defined by
\begin{eqnarray}
g(x,v)=v_\mu\gamma^\mu,
\label{tran}
\end{eqnarray}
 where $x\in M$, $v$ is
unit vector of the $2n$ dimensional vector space, which is
the element of the fiber of $S(H)$ at the point $x$,
and $\gamma^\mu$ is the $SO(2n)$ gamma matrices restricted
on the space of positive chirality spinors.
The transition function is interpreted as the tachyon field
created by the open string stretched between the D$(2n+p)$-brane
and the anti D$(2n+p)$-brane, and this tachyon configuration
(\ref{tran})
induces a unit D$p$-brane charge \cite{Wi}.
Therefore, this configuration should be physically
identified with D$p$-brane world-volume characterized
by $(M,E,\varphi)$. This is the physical
meaning of the equivalence relation (c).

As mentioned above,
one can show that the topological K-homology is
isomorphic to the analytic K-homology, which we described in
the previous subsection. The isomorphism
\begin{eqnarray}
\mu_i: K^{top}_i(X) \stackrel{\sim}{\rightarrow} K_i(X)~~~(i=0,1)
\label{isom}
\end{eqnarray}
is given as follows.
To be specific, we will explain the $i=1$ case.
Let $(M,E,\varphi)$ be an element of $K^{top}_1(X)$.
Since $M$ is an odd dimensional closed Spin${}^c$ manifold,
we can define a spin bundle $S$ associated to
the spinor representation of the Spin${}^c$ group.
$\Gamma(S\otimes E)$ denotes the space of smooth sections of the
vector bundle $S\otimes E$ on $M$.
We can define a Dirac operator $D$ on $\Gamma(S\otimes E)$ by
choosing a connection on the bundle $S\otimes E$ as usual.
$\Gamma(S\otimes E)$ is equipped with an inner product,
and the Hilbert space $\cH=L^2(M,S\otimes E)$ is defined
by the completion of $\Gamma(S\otimes E)$ with respect to
the inner product.
The Dirac operator $D$ can be thought of as an (unbounded) operator
on $\cH$.
The representation $\phi:C(X)\rightarrow \B(\cH)$ is defined by
the multiplication of the function
$\phi(f)\equiv f\circ\varphi$ for each function $f\in C(X)$.
Thus, we have obtained an unbounded Fredholm module
 $(\cH,\phi,D)$. It can be shown that
this defines an element of the K-homology $K_1(X)$
irrespective of the choice of the connection on $S\otimes E$,
and furthermore, it gives a well-defined
map from $K_1^{top}(X)$ to $K_1(X)$, which
turns out to be an isomorphism.

As we have seen in section \ref{CSandD},
the tachyon operator for the
 D-brane solution in \cite{Te}
is nothing but a Dirac operator acting
 on a spin bundle on $M=\R^{2m+1}$.
Therefore this D-brane configuration in the K-matrix theory
corresponds to the element of topological K-homology
with $M=\R^{2m+1}$ and $E=I$ (trivial line bundle),
which is interpreted as a D$(2m)$-brane extending along $\R^{2m+1}$
with trivial Chan-Paton bundle. This is exactly
what we expect from the calculation of
the tension and the Chern-Simons term \cite{Te}.
The isomorphism (\ref{isom}) suggests that
we can always obtain a world-volume interpretation for
each configuration in the K-matrix theory.
Since we have a clear geometrical interpretation of the
D-brane configurations in the topological
K-homology, it provides a convincing evidence for
our proposal that the D-brane configurations
in the K-matrix theory is classified by the (analytic) K-homology.

\subsection{Chern character and Chern-Simons terms}

In the K-theory description of D-branes,
a D-brane is constructed as the gauge configuration
on non-BPS D9-brane system (type IIA)
or D9-\AD9 system (type IIB),
and the D-brane charge is classified by the K-theory groups
$K^*(X)$ \cite{Wi,Ho}.
But when we are not interested in the torsion part of the K-theory groups,
it is enough to use the cohomology group $H^*(X)$ to
classify the D-brane charges, and we can read them
from the Chern-Simons terms.

In order to write down the Chern-Simons term,
the Chern character plays a crucial role \cite{MiMo}.
Namely, the Chern character induces isomorphisms
\begin{eqnarray}
\ch&:& K^0(X)\otimes\Q\stackrel{\sim}{\rightarrow} H^{even}(X;\Q),
\label{KandH1}\\
\ch&:& K^1(X)\otimes\Q\stackrel{\sim}{\rightarrow} H^{odd}(X;\Q),
\label{KandH2}
\end{eqnarray}
between the K-theory groups tensored by $\Q$ and the cohomology groups.
Using these maps
the Chern-Simons term for the world-volume
theory of D9-branes (in the flat background)
can be written as
\begin{eqnarray}
S_{\rm CS}=\int_X C \wedge\ch(x),
\label{CS1}
\end{eqnarray}
where $C$ is the formal sum of RR-fields and $x$ is an element
of $K^*(X)$. The Chern character
$\ch(x)$ can be explicitly written in terms of superconnections
\cite{Qu}.
For example, in type IIA string theory,
the relevant superconnection is defined by
\begin{eqnarray}
i {\cal A} &=& \left(
\begin{array}{cc}
i A & T  \\
T & i A
\end{array}
\right)= i A \otimes I_2
+T \otimes \sigma^1, \\
{\cal F}&=& d {\cal A}-i {\cal A}^2
=i \mat{F-T^2, D T,DT,F-T^2},
\end{eqnarray}
where $T$ is the tachyon and $A$ is the gauge field
on the world-volume of the non-BPS D9-brane.
Using these variables, (\ref{CS1}) becomes \cite{KrLa,TaTeUe}
\begin{eqnarray}
S_{\rm CS}
&=&\int_X C\wedge \Tr_{ N} \left( \Tr_2
\left( \sigma^1 e^{{\cal F}} \right) \right).
\end{eqnarray}
In this formula, one can easily show that
 $\Tr_{ N} \Tr_2 \sigma^1 e^{\cal F}$ is a closed form,
which ensures the invariance under the gauge transformation
$C\rightarrow C+d\Lambda$.

What is the counterpart of this in the K-homology?
There is a theorem analogous to (\ref{KandH1}) and (\ref{KandH2}),
which claims that the K-homology group is isomorphic to the
ordinary homology group if it is tensored by $\Q$ \cite{BaDo}:
\begin{eqnarray}
\ch.&:& K_0(X)\otimes\Q\stackrel{\sim}{\rightarrow} H_{even}(X;\Q),
\label{Chhomo1}\\
\ch.&:& K_1(X)\otimes\Q\stackrel{\sim}{\rightarrow} H_{odd}(X;\Q).
\label{Chhomo2}
\end{eqnarray}
Here the element $(M,E,\varphi)\in K_*(X)$ of topological K-homology
is mapped to
\begin{eqnarray}
\ch.(M,E,\varphi)=\varphi_*(\ch(E)\cup\Td(TM)\cap[M]).
\label{Ch}
\end{eqnarray}
Since the homology group in (\ref{Chhomo1}) and (\ref{Chhomo2})
classifies the world-volume cycle of
the D-brane which corresponds to the element $(M,E,\varphi)$ of
the K-homology, the Chern-Simons term should be
written by integrating $C$ over the cycle (\ref{Ch}).
Hence we obtain
\begin{eqnarray}
S_{\rm CS}=\int_M \varphi^*C\wedge\ch(E)\wedge\Td(TM),
\end{eqnarray}
which agrees with the Chern-Simons term
for a D-brane of world-volume $M$ with Chan-Paton
bundle $E$, dropping the factor that comes from
the background curvature on $X$ \cite{AlSc,MiMo}.
This is again consistent with the interpretation that
$M$ is the world-volume of the brane and $E$ is the Chan-Paton
bundle on it.

On the other hand, we have an analytic description of the
K-homology group. How can we define the Chern character
in this formulation and relate it to the Chern-Simons terms?
To answer this question,
note that we can rewrite the Chern-Simons term for the IIA
K-matrix theory (\ref{CSterm2}) as
\beq
S_{\rm CS}= \sum_n \Psi_{2n+1}
\left( C_{\mu_1\cdots\mu_n}, x^{\mu_1},
\cdots, x^{\mu_n} \right),
\eeq
where
\beqa
&&\Psi_{2n+1} (a^0,a^1,\cdots,a^{2n+1})
= \int_{\sum s_i=1, s_i \geq 0} ds_0 \cdots d s_{2n} \times \nonumber \\
&&\hspace{2.1cm}
\Tr_{\cH} \left( \phi(a^0) e^{-s_0 T^2} [T,\phi(a^1)] e^{-s_1 T^2}
\cdots [T,\phi(a^{2n+1})] e^{-s_{2n+1} T^2}
\right),
\label{JLO}
\eeqa
and $a^i\in\cA$.
$\Psi=(\Psi_{2n+1})$ is known as (odd) JLO cocycle \cite{JLO}
associated with the unbounded Fredholm module $(\cH,\phi,T)$.
This is a Chern character of the K-homology $K^1(\cA)$
that takes value in the entire cyclic cohomology
$HE^*(\cA)$. (See \cite{Connes}.)
In this case, the JLO cocycle has similar role as
the usual Chern character or superconnection.
The JLO formula (\ref{JLO}) can also be used in the case that
$\cA$ is noncommutative.
It seems to be quite natural that the CS-term for the D-brane
in noncommutative manifold is also obtained from the Chern
character of the K-homology.
However, it is not clear in the above formula
how to incorporate the Myers' terms \cite{My}
and the RR-fields which are not restricted to be constant.

\subsection{KK-theory}
\label{KK-theory}

In this subsection, we will leave the K-matrix theory for a while
and deal with the classification of D-branes in a slightly more
general context.
Let us consider the field theory of higher dimensional D-branes
as a fundamental theory instead of D-instantons,
to be precise, the non-BPS D$p$-brane system
or D$p$-\AD$p$ system in type II string theory.
As for the K-matrix theory, we can classify possible stable
configurations
of the theory and claim that in this case the appropriate
group is KK-theory,
which is a generalization of both K-theory and K-homology.
We will also see that this classification is in fact equivalent to
that of K-matrix theory in a simple example.

Let $(A,B)$ be a pair of $C^*$-algebras.
\footnote{Here we assume that both $A$ and $B$ are unital.}
The KK-group $KK(A,B)$ is an Abelian group associated with $(A,B)$,
which is covariant in $B$ and contravariant in $A$.
Roughly speaking, it is defined by equivalence classes of
Hilbert $(A,B)$-bimodules, called Kasparov modules.
\footnote{For notations used here and mathematical details,
see \cite{Bl,Connes}.}
There is a natural Abelian group structure
on $KK(A,B)$ induced by the direct sum of Kasparov modules.
More generally, $KK^n(A,B)$ is defined by $KK(A,B{\otimes}\C_n)$,
where $\C_n$ is the complex Clifford algebra of $\R^n$.

Although there are various expressions for Kasparov modules,
the Fredholm picture of $KK(A,B)$,
in which the tachyon operator $T$ carries almost all non-trivial
information, is the appropriate one for us.

In precise, an odd Kasparov module in the Fredholm picture
is defined by a triple $(\cH_{B},\phi,T)$, where
\begin{itemize}
\item $\cH_{B}=B^\infty$ is a Hilbert space over $B$,
\item $\phi: A\rightarrow\B(\cH_{B})$ is a *-homomorphism,
\item $T$ is a self-adjoint operator in $\B(\cH_{B})$ such that
\begin{eqnarray}
T^2-1,~[T,\phi(a)]~\in\K(\cH_{B})=B\otimes \K~~\mbox{for}~
{}^\forall a\in A.
\end{eqnarray}
\end{itemize}
Note the similarity in the case of K-homology.
Since $\cH_{B}$ is a family of Hilbert spaces on the
(possibly noncommutative) space $B$,
a triple $(\cH_{B},\phi,T)$ is also a family of Fredholm
modules on the space $B$.
Of course, for $B=\C$ (one point space) it
reduces to the (odd) Fredholm module described in section \ref{analyK}.
The elements of $KK^1(A,B)$ are homotopy equivalence classes
of odd Kasparov modules,
with the similar equivalence relations
as those used in the definition of K-homology
in section \ref{analyK}.

An even Kasparov module
$(\cH_{B}^{(0)}\oplus \cH_{B}^{(1)},\phi^{(0)}\oplus \phi^{(1)},\wh F)$
is defined by almost the same condition above,
except that it has $\Z_2$ grading given by a standard self-adjoint
involution operator $\gamma$.
Namely, in the matrix form one has
\begin{equation}
\wh \cH_{B}=\left(
\begin{array}{c}
  \cH_{B}^{(0)} \\
  \cH_{B}^{(1)}
\end{array}
\right), ~~~
\gamma=\matt[1,0,0,-1], ~~~
\wh \phi=\matt[\phi^{(0)},0,0,\phi^{(1)}],~~~
\wh F=\matt[0,T^*,T,0].
\end{equation}
The equivalence classes of even Kasparov modules define
$KK(A,B)=KK^0(A,B)$.
Note that odd Kasparov modules are also described by the matrix form
as above with $\phi^{(0)}=\phi^{(1)}$ and $T=T^*$, which is equivalent to
$\phi=I_2\otimes \phi^{(0)}$ and $F=\sigma^1\otimes T$, using generators
$\{I_2,\sigma^1\}$ of $\C_1$. In fact, one can show that
$KK^1(A,B)$ is isomorphic to $KK(A,B\otimes \C_1)$.

This set-up fits nicely to the classification of D-branes
made from the non-BPS D$p$-brane or
D$p$-\AD$p$ system with $p+1$ dimensional world-volume $B$.
Stable D-brane configurations made out of the non-BPS D$p$-brane
and embedded in transverse space $A$ are classified by
$KK^1(A,B)$, and that of D$p$-\AD$p$ system are classified by
$KK^0(A,B)$.
This picture is clearer when the whole space-time is a product
space $A\otimes B$. For the sake of simplicity,
let us explain the physical interpretation of $KK^1(A,B)$
using the non-BPS D$p$-brane system.
Let $C(X)=C(N)\otimes C(M)$ be a
$10$ dimensional space-time, where
$B=C(M)$ is a $p+1$ dimensional world-volume of non-BPS D$p$-branes
with coordinates $x^\alpha \ (\alpha=0,1,\cdots ,p)$,
and $A=C(N)$ is a $9-p$ dimensional space transverse to $M$
with coordinates $y^i \ (i=p+1,\cdots,9)$.
Then $\cH_{B}$ is the Chan-Paton bundle on $M$ with
(infinite) dimensional Hilbert spaces as fibers.
The *-homomorphism $\phi$ is given by $\phi(y^i)=\Phi^i(x)$
where $\Phi^i(x)$ are the $9-p$ scalar fields
on the non-BPS D$p$-branes wrapped on $M$,
and the operator $T=T_b(x)$ is  the (normalized) tachyon field.
They fit into odd Kasparov modules in the Fredholm picture
and lead $KK^1(A,B)$.
$KK^0(A,B)$ is quite analogous. In this case $\Z_2$ grading
corresponds to Chan-Paton indices of D$p$-branes and \AD$p$-branes.
This construction generalizes
the K-theory classification of D-brane charges
as well as the K-homology classification of D-brane configurations
given in section \ref{analyK}.
In particular, it agrees with the Fredholm picture of
K-theory used in \cite{HaMo,Wi2} for the $p=9$ case.

Note that a configuration can be expanded in $i$-th direction
and localized in $\alpha$-th direction at the same time.
For example, if a tachyon field is roughly given by
$T\sim p_i\gamma^i+x^\alpha\gamma_\alpha$, it represents such a configuration.

In the Fredholm picture above we do not care about gauge fields,
since any bundle of infinite dimensional
separable Hilbert spaces is known to be trivial and the tachyon field
carries the topological information
instead of it.
There is, however, another interesting description of
Kasparov modules, the unbounded version of the Fredholm picture,
which is the analog of the spectral-triple description for K-matrix theory.
In this picture gauge fields can be incorporated through
superconnection $Z=1\otimes \nabla+\sigma^1\otimes T$ \cite{Connes}.

In summary, we claim that the classification based on
non-BPS D$p$-brane system corresponds to $KK^1(A,B)$ and that based on
D$p$-\AD$p$ system corresponds to $KK(A,B)$.

We summarize basic properties for KK-theory
\cite{Bl,Connes}:
\begin{itemize}
\item $KK(A,B)$ includes both (algebraic) K-theory
and (analytic) K-homology:
\begin{eqnarray}
\label{eq:KK}
KK(\C,B)=K_0(B), \quad KK(A,\C)=K^0(A).
\end{eqnarray}
\item There is a bilinear associative intersection product,
called Kasparov product: for any $C^*$-algebras $A_1,A_2,B_1,B_2$,
\begin{equation}
KK(A_1, B_1\otimes D)\otimes _{D}
KK(D\otimes A_2,B_2) \rightarrow KK(A_1\otimes A_2,B_1\otimes B_2),
\end{equation}
which is essentially given by the inner tensor product of two bimodules.
This especially makes $KK(A,A)$ a ring with unit element $1_A$.
\item Periodicity: for $n$ even,
\begin{equation}
KK^{n}(A,B):=KK(A,B\otimes \C_n)\simeq KK(A,B),
\end{equation}
so only $KK(A,B)$ and $KK^1(A,B)$ are independent.
\item Duality: assume that there are two elements
$\alpha\in KK(A\otimes B,\C), \ \beta\in KK(\C,A\otimes B)$ such that
$\beta\otimes _{A}\alpha=1_B\in KK(B,B), \ \beta\otimes _{B}\alpha=1_A\in KK(A,A)$.
Then it follows that for any pair $(D,E)$ of $C^*$-algebras the maps
\begin{eqnarray}
\label{eq:K-dual}
&& \otimes _{A}\alpha: \ KK(D,A\otimes E)\rightarrow KK(D\otimes B,E), \\
&& \otimes _{B}\alpha: \ KK(D,B\otimes E)\rightarrow KK(D\otimes A,E)
\end{eqnarray}
are isomorphisms
(with inverse $\beta\otimes_{B}$ and $\beta\otimes_{A}$, respectively).
Such a pair $A$ and $B$ are called K-dual of each other.
In particular, for $D=E=\C$ it gives Poincare duality between K-theory
and K-homology:
\begin{equation}
\label{eq:Poincare}
K_*(A)\simeq K^*(B), \ K_*(B)\simeq K^*(A).
\end{equation}
A simple example of this is given for $A=B=C(M)$,
where $M$ is a compact Spin${}^c$ manifold.
In this case,
$\alpha\in KK^n(C(M)\otimes C(M),\C) \ (n=\mbox{dim}M\,\,\mbox{mod } 2)$
is known as the Dirac K-cycle $[M]$ and gives
isomorphism (\ref{eq:K-dual}).
\end{itemize}

Since non-BPS D$p$-branes or D$p$-\AD$p$ system
considered above are also (unstable) configurations constructed
in the K-matrix theory, the classification of
D-branes built from non-BPS D$p$-branes or D$p$-\AD$p$ system
should be related to that built from the K-matrix theory.
This is shown by the isomorphism (\ref{eq:K-dual}).
For example, take the Dirac K-cycle
$\alpha\in KK^n(C(M)\otimes C(M),\C)$
with $n=\mbox{dim}\,M$ (mod 2) and set $D=C(N), E=\C$.
Then, it gives the following isomorphism
\begin{equation}
KK^{i+n}(C(N),C(M))\simeq KK^{i}(C(N\times M),\C),
\label{KKequiv}
\end{equation}
where $i+n$ and $i$ are understood mod 2.
This isomorphism (\ref{KKequiv}) generalizes
the isomorphism between K-theory and K-homology
(\ref{Kisom}).
For the case $i=0$, the right hand side of
(\ref{KKequiv}) is the group which classifies
the D-branes in IIB K-matrix theory.
On the other hand, the left hand side is
the group which classifies the D-branes made
from non-BPS D$p$-branes (for $n=$ odd) or D$p$-\AD$p$ system
(for $n=$ even) wrapped on $M$.
Analogous relations for type IIA string theory holds.
This means that various descriptions
give the same result as expected.

The KK-theory unifies various constructions
of D-branes via tachyon condensation.
It would be useful to analyze some
duality transformations, such as T-duality.
We will come back to this issue elsewhere.

\section{Boundary states and spectral triples}
\label{BdryState}

Given a configuration $(\cH,\{\Phi^\mu\},T)$ in
the K-matrix theory, we can construct, at least formally,
a (off-shell) boundary state that corresponds to it.
There are many applications for this approach.
The boundary state can be used to evaluate the BSFT action
and calculate the tension and RR-charge of the D-brane
represented by the spectral triple.
Moreover, we can explicitly see how the spectral triples
represent higher dimensional D-branes
in an analogous way that is
given in \cite{Is,Ok} for bosonic string theory.
This section is devoted to explain these results and
give another viewpoint for
 our proposal that D-branes are represented by spectral
triples.
To be specific, we will examine the boundary states for
 type IIA non-BPS D-instantons, though the generalization
to the other cases is straightforward.

\subsection{Boundary states}

In this section, we review in some detail the boundary state approach
for D-branes.

The boundary state ( See \cite{DiLi} for a review.)
of the non-BPS D-instanton in type IIA string theory
is given by
\begin{eqnarray}
\ket{\wh D(-1)}=\frac{1}{\sq}\left(
\ket{D;+}_{\rm NS}-\ket{D;-}_{\rm NS}\right).
\label{Dstate}
\end{eqnarray}
Here $\ket{D;\pm}_{\rm NS}$ are boundary states which satisfy
Dirichlet boundary condition for all directions
and NS-fermions with $\pm$ spin structure.
They can be expressed as
\footnote{We will omit the ghost part of the boundary states,
which do not play an important role in our analysis.}
\begin{eqnarray}
\ket{D;\pm}_{\rm NS}&=&\ket{x=0}
\ket{\theta=0;\pm}_{\rm NS},
\end{eqnarray}
using  coherent states
\begin{eqnarray}
\ket{x}&=&
\exp\left(\sum_{m=1}^\infty\left(-\half x_{-m} x_m-a^\dag_m\wt a^\dag_m
+a^\dag_m x_m+x_{-m} \wt a^\dag_m\right) \right)\ket{x_0},\\
\ket{\theta;\pm}_{\rm NS}
&=&
\exp\left(\sum_{r>0}\left(
-\half\theta_{-r}\theta_r \pm i\psi^\dag_r\wt\psi^\dag_r
+\psi^\dag_r\theta_r\mp i\theta_{-r}\wt\psi^\dag_r
\right)\right)\ket{0}_{\rm NS},
\end{eqnarray}
where
\begin{eqnarray}
a_m^\mu=i\alpha_m^\mu/\sqrt{m},~~~
a_{-m}^\mu=a_m^{\mu\dag}=-i\alpha_{-m}^\mu/\sqrt{m},~~~(m>0).
\end{eqnarray}
These are eigen states of operators $X^\mu(\sigma)$ and
$\Theta^\mu_\pm(\sigma)$ defined as
\begin{eqnarray}
X^\mu(\sigma)&=&\wh x_0^\mu+\sum_{m\ne 0}\frac{1}{\sqrt{|m|}}
(a_m^\mu e^{-im\sigma}+\wt a_m^\mu e^{im\sigma}),\\
\Theta^\mu_\pm(\sigma)&=&
\psi^\mu(\sigma) \pm i\wt\psi^\mu(\sigma)
=\sum_{r}(\psi^\mu_r e^{-ir\sigma}\pm i\wt\psi^\mu_r e^{ir\sigma}),
\end{eqnarray}
and satisfy
\begin{eqnarray}
X^\mu(\sigma)\ket{x}&=&x^\mu(\sigma)\ket{x},
\label{coherent}\\
\Theta_\pm^\mu(\sigma)\ket{\theta;\pm}&=&
\theta^\mu(\sigma)\ket{\theta;\pm},
\end{eqnarray}
\begin{eqnarray}
\int [dx]\ket{x}\bra{x} &=&1,\\
\int [d\theta]
\ket{\theta;\pm}\bra{\theta;\pm} &=&1,\
\end{eqnarray}
where
\begin{eqnarray}
x^\mu(\sigma)=x_0^\mu+\sum_{m\ne 0}\frac{1}{\sqrt{|m|}}\,
x_m^\mu e^{-im\sigma},~~~
\theta^\mu(\sigma)
=\sum_{r}\theta_r^\mu e^{-ir\sigma}.
\end{eqnarray}

The boundary state for the Neumann boundary condition is
obtained by integrating these coherent states as
\begin{eqnarray}
\ket{N;\pm}_{\rm NS}&=&\int[dx][d\theta]\ket{x}\ket{\theta;\pm}_{\rm NS}\\
&=&e^{+\sum_{m=1}^\infty a^\dag_m\wt a^\dag_m
 \mp \sum_{r>0} i\psi^\dag_r\wt\psi^\dag_r}\ket{0}_{\rm NS}.
\end{eqnarray}
Here the ground state is the zero momentum state
$\int \frac{dx_0}{\sqrt{2\pi}}\ket{x_0}=\ket{0}$.

In fact, one can easily check that this state satisfies
the Neumann boundary condition,
\begin{eqnarray}
0&=&\int[dx][d\theta]\frac{\delta}{i\delta x^\mu(\sigma)}
\ket{x}\ket{\theta;\pm}\\
&=&\int[dx][d\theta]\frac{\delta}{i\delta x^\mu(\sigma)}
e^{i\int d\sigma' P_\nu(\sigma')\,x^\nu(\sigma')}
\ket{x=0}\ket{\theta;\pm}\\
&=&P_\mu(\sigma)\ket{N;\pm},
\end{eqnarray}
where  $P_\mu(\sigma)$ is the momentum operator conjugate to $X^\mu$.
\begin{eqnarray}
P_\mu(\sigma)&=&\half \sum_{m=-\infty}^\infty\left(
\alpha_{m\mu}\, e^{-im\sigma}+\wt\alpha_{m\mu} \,e^{im\sigma}
\right).
\end{eqnarray}

Therefore the boundary states for a D$p$-brane stretched
along $x^\alpha$ ($\alpha=0,1,\dots,p$) axes
and located at $x^i=0$ ($i=p+1,\dots,9$)
are linear combinations of the following states.
\begin{eqnarray}
\ket{Bp;\pm}=\int[dx^\alpha][d\theta^\alpha]
\ket{x^\alpha, x^i=0}\ket{\theta^\alpha, \theta^i=0;\pm}.
\label{Bp}
\end{eqnarray}

We can similarly construct the boundary states in RR-sector.
In this case, however, we should be careful about the fermion zero mode.
The coherent state for the RR-fermion operator $\Theta_\pm(\sigma)$ is
\begin{eqnarray}
\ket{\theta;\pm}_{\rm RR}
&=&
\exp\left(\sum_{n=1}^\infty\left(
-\half\theta_{-n}\theta_n \pm i\psi^\dag_n\wt\psi^\dag_n
+\psi^\dag_n\theta_n\mp i\theta_{-n}\wt\psi^\dag_n
\right)\right)\, e^{i\half(\psi_0\mp i\wt\psi_0)\theta_0}
\ket{D;\pm}^{(0)}_{\rm RR},
\nonumber\\
\end{eqnarray}
where $\ket{D;\pm}^{(0)}_{\rm RR}$ is the ground state
defined as follows.

 Let us define the zero mode operators
\begin{eqnarray}
\psi_\pm^\mu=\frac{1}{\sqrt{2}}
(\psi_0^\mu\pm i\wt\psi_0^\mu),
\end{eqnarray}
which satisfy the following anti-commutation relations
\begin{eqnarray}
\{\psi_+^\mu,\psi_-^\nu\}=\delta^{\mu\nu},
~~~\{\psi_+^\mu,\psi_+^\nu\}=\{\psi_-^\mu,\psi_-^\nu\}=0.
\end{eqnarray}
These are nothing but the anti-commutation relations of
$SO(10,10)$ gamma matrices which we encountered in (\ref{1010}).
So we can regard $\psi_\pm$ as $SO(10,10)$ gamma matrices.
$\ket{D;\pm}^{(0)}_{\rm RR}$ is one of the states which
belong to the irreducible representation of the gamma matrices,
and satisfies the the Dirichlet boundary condition
\begin{eqnarray}
\psi_\pm^\mu\ket{D;\pm}^{(0)}_{\rm RR}=0.
\end{eqnarray}
Similarly, the Neumann boundary condition implies
\begin{eqnarray}
\psi_\mp^\mu\ket{N;\pm}^{(0)}_{\rm RR}=0,
\end{eqnarray}
and hence the ground state with a mixed boundary condition
as (\ref{Bp}) should satisfy
\begin{eqnarray}
\psi_\mp^\alpha\ket{Bp;\pm}^{(0)}_{\rm RR}&=&0~~~
(\alpha=0,1,\dots,p),\\
\psi_\pm^i\ket{Bp;\pm}^{(0)}_{\rm RR}&=&0~~~
(i=p+1,\dots,9).
\end{eqnarray}
They are constructed by acting $\psi_-^\mu$ on $\ket{D;+}^{(0)}_{\rm RR}$ as
\begin{eqnarray}
\ket{Bp;+}^{(0)}_{\rm RR}&=&
\prod_{\alpha=0}^p\psi_-^\alpha\ket{D;+}^{(0)}_{\rm RR},
\label{zero}\\
\ket{Bp;-}^{(0)}_{\rm RR}&=&
\prod_{\alpha=0}^p\psi_+^\alpha
\prod_{i=p+1}^9\psi_-^i
\ket{Bp;+}^{(0)}_{\rm RR},
\end{eqnarray}
up to phase factor. We fix the phase difference among
$\ket{Bp;\pm}^{(0)}_{\rm RR}$ by these relations.

Then one can easily show that they satisfy
\begin{eqnarray}
(-1)^F\ket{Bp;\pm}_{\rm NS}&=&-\ket{Bp;\mp}_{\rm NS},
\label{F1}\\
(-1)^{\wt F}\ket{Bp;\pm}_{\rm NS}&=&-\ket{Bp;\mp}_{\rm NS},
\label{F2}\\
(-1)^F\ket{Bp;\pm}_{\rm RR}&=&\ket{Bp;\mp}_{\rm RR},\\
(-1)^{\wt F}\ket{Bp;\pm}_{\rm RR}&=&
(-1)^{p+1}\ket{Bp;\mp}_{\rm RR},
\end{eqnarray}
where $F$ ($\wt F$) is the world-sheet left (right)
moving fermion number operator.
Note that $(-1)^{F}$ and $(-1)^{\wt F}$ act as
\begin{eqnarray}
(-1)^{F} &=& \prod_{\mu=0}^p(\psi_+^\mu+\psi_-^\mu),\\
(-1)^{\wt F}&=& \prod_{\mu=0}^p(\psi_+^\mu-\psi_-^\mu)
\end{eqnarray}
on the RR ground states.

The linear combinations that survive after
GSO projection are
the boundary states of non-BPS D$p$-branes
with odd (even) $p$
\begin{eqnarray}
\ket{\wh Dp}=\frac{1}{\sqrt{2}}\left(
\ket{Bp;+}_{\rm NS}-\ket{Bp;-}_{\rm NS}\right),
\end{eqnarray}
and BPS D$p$-branes with even (odd) $p$
\begin{eqnarray}
\ket{Dp}&=&\half
\left(\ket{Bp;+}_{\rm NS}-\ket{Bp;-}_{\rm NS}
+\ket{Bp;+}_{\rm RR}+\ket{Bp;-}_{\rm RR}\right),
\label{BPS}
\end{eqnarray}
in type IIA (IIB) string theory.

When we turn on the tachyon fields, gauge fields and so on,
the boundary states $\ket{Bp;\pm}$ are modified as
\begin{eqnarray}
\ket{Bp;\pm}_{S_b}&=&
\int[dx^\alpha][d\theta^\alpha]
\,e^{-S_b (x,\theta)}
\ket{x^\alpha, x^i=0}\ket{\theta^\alpha, \theta^i=0;\pm}\\
&=&
e^{-S_b (X,\Theta_{\pm})}\ket{Bp;\pm},
\end{eqnarray}
where $S_b (X,\Theta)$ is the boundary interaction.
The boundary interaction for
non-BPS D9-branes in type IIA string theory is given
in \cite{KuMaMo2,KrLa,TaTeUe} as
\begin{eqnarray}
e^{-S_b(X,\Theta)}=\int[d\eta^I]
\exp\left\{
\oint d\sigma\left(
\frac{1}{4}\dot\eta^I\eta^I+
\sum_{k=0}^{2m} \frac{1}{2k!}(M_1-M_0^2)^{I_1\cdots I_k}
\eta^{I_1}\cdots\eta^{I_k}
\right)\right\},
\label{bdry}
\end{eqnarray}
where $\dot\eta=\del_\sigma\eta$ and
\begin{eqnarray}
M_0&=&\mat{iA_\mu(X)\,\Theta^\mu,T(X),T(X),
iA_\mu(X)\,\Theta^\mu},\\
M_1&=&\mat{iA_\mu(X) \dot X^\mu
+i\del_\nu A_\mu(X)\,\Theta^\nu\Theta^\mu,
\del_\mu T(X)\,\Theta^\mu, \del_\mu T(X)\,\Theta^\mu,
iA_\mu(X)\dot X^\mu+i\del_\nu A_\mu(X)\,\Theta^\nu\Theta^\mu},
\end{eqnarray}
\begin{eqnarray}
M_1-M_0^2=
\mat{iA_\mu\dot X^\mu-T^2+ \frac{i}{2}F_{\nu\mu}\,\Theta^\nu\Theta^\mu,
D_\mu T\,\Theta^\mu,D_\mu T\,\Theta^\mu,
iA_\mu\dot X^\mu-T^2+ \frac{i}{2}F_{\nu\mu}\,\Theta^\nu\Theta^\mu}.
\end{eqnarray}
Here we assume that $M_0$ and $M_1$ are $2^m\times 2^m$ matrices,
and $M^{I_1\cdots I_k}$ stand for
coefficients with respect to the following expansion of
a $2^m\times 2^m$ matrix $M$
\begin{eqnarray}
M=\sum_{k=0}^{2m} \frac{1}{2k!}M^{I_1\cdots I_k}
\gamma^{I_1\cdots I_k},
\end{eqnarray}
where $\gamma^{I_1\cdots I_k}$ are the
skew-symmetric products of $SO(2m)$ gamma matrices
$\gamma^I$.

The boundary interaction for non-BPS D-instantons
can be obtained by replacing
\begin{eqnarray}
A_\mu&\rightarrow& \Phi_\mu,\\
F_{\nu\mu}&\rightarrow& i[\Phi_\nu,\Phi_\mu],\\
D_\mu T&\rightarrow& i[\Phi_\mu,T],\\
\dot X^\mu &\rightarrow& 2 P^\mu,\\
 \Theta^\mu_\pm &\rightarrow&\Theta^\mu_\mp=:-2i\Pi^\mu_\pm,
\end{eqnarray}
where $P^\mu$ and $\Pi_\pm^\mu$ are canonical momentum
conjugate to $X^\mu$ and $\Theta_\pm^\mu$ respectively.
Thus we obtain the boundary interaction (\ref{bdry}) with
\begin{eqnarray}
M_1-M_0^2=
\mat{i\Phi_\mu P^\mu-T^2-\half{[\Phi_\nu,\Phi_\mu]}\,\Pi^\nu\Pi^\mu,
{[\Phi_\mu, T]}\,\Pi^\mu,{[\Phi_\mu, T]}\,\Pi^\mu,
i\Phi_\mu P^\mu-T^2-\half{[\Phi_\nu,\Phi_\mu]}\,\Pi^\nu\Pi^\mu}.
\label{M}
\end{eqnarray}
Here we rescaled $2\Phi_\mu\rightarrow \Phi_\mu$ to
avoid factor 2 appearing everywhere.

In the case that $\Phi_\mu$ and $T$ are operators acting on an
infinite dimensional Hilbert space
$\cH$, we cannot assume that $M_0$ and $M_1$ are  $2^m\times 2^m$ matrices.
Instead, we treat these as $2\times 2$ matrices with operator
coefficients. Namely, expanding $M_1-M_0^2$ with respect to
$I_2=\left({1~~\atop~~1}\right)$ and
$\sigma^1=\left({~~1\atop1~~}\right)$ as
\begin{eqnarray}
M_1-M_0^2 =
\left(i\Phi_\mu P^\mu-T^2-\half{[\Phi_\nu,\Phi_\mu]}\,\Pi^\nu\Pi^\mu\right){I_2}
+\left({[\Phi_\mu T]}\,\Pi^\mu\right)\,{\sigma^1},
\end{eqnarray}
we obtain the boundary interaction
\begin{eqnarray}
e^{-S_b(P,\Pi)}=\int[d\eta]\,
\Tr_{\cH}{\rm P}
\exp\left\{
\oint d\sigma\left(
\frac{1}{4}\dot\eta\eta+
i\Phi_\mu P^\mu-T^2-\half{[\Phi_\nu,\Phi_\mu]}\,\Pi^\nu\Pi^\mu
+{[\Phi_\mu T]}\,\Pi^\mu\eta
\right)\right\}.\nonumber\\
\label{bdry2}
\end{eqnarray}
In this equation,
we should take the path ordered trace $\Tr_{\cH}{\rm P}$
on the Hilbert space $\cH$, which corresponds to integrating out
the rest of $\eta^I$ (except for $\eta$)
in the formula (\ref{bdry}).

Collecting all these together,
the boundary state for non-BPS D-instantons with the
boundary interaction is given as
\begin{eqnarray}
\ket{\wh D(-1)}_{S_b}&=&
P\wt P_+ \,e^{-S_b(P,\Pi_+)}\ket{D;+}_{\rm NS}+
P\wt P_- \,e^{-S_b(P,\Pi_+)}\ket{D;+}_{\rm RR},
\label{Bint}
\end{eqnarray}
where $P=\half(1+(-1)^F)$ and  $\wt P_\pm=\half(1\pm(-1)^{\wt F})$
are GSO projection operators.

It is important to note that unlike the non-BPS D-instanton
state without boundary interaction (\ref{Dstate}),
 the RR-sector may not be projected out,
since the boundary interaction carries the fermion zero modes.
Let us check that the boundary state
(\ref{Bint}) becomes (\ref{Dstate}) when the boundary
interaction is turned off, i.e. $T=\Phi_\mu=0$.
In this case, the boundary interaction for RR-sector vanishes
since the zero mode of $\eta$ is not saturated in the $\eta$ integral.
The boundary interaction for NS-sector becomes
\begin{eqnarray}
e^{-S_b(P,\Pi)}/(\Tr_\cH\,1)&=&
\int[d\eta]\,
e^{ \oint d\sigma\left(\frac{1}{4}\dot\eta\eta\right)}
=\int\prod_{r>0}d\eta_r d\eta_{-r}\,
e^{\left(\pi r\eta_r \eta_{-r}\right)}\\
&=&\prod_{r=1/2,1/3,\dots}\pi r\\
&=&\sqrt{2}.
\label{root2}
\end{eqnarray}
Here we used $\zeta$-function regularization
(see Appendix \ref{zeta}) in the last step.
\footnote{This formula is obtained by indirect argument
 in \cite{Wi}
and used to explain the tension of non-BPS D-branes.}
Note that the factor $(\Tr_\cH\,1)$ represents that
we have infinite number of non-BPS D-instantons
when $T=0$.
Then, using (\ref{root2}), (\ref{F1}) and (\ref{F2}),
(\ref{Dstate}) is reproduced from (\ref{Bint}).

\subsection{Tachyon condensation}

In the last subsection, we constructed the boundary state (\ref{Bint})
with the boundary interaction (\ref{bdry2}),
which corresponds to the D-brane represented by the
 configuration $(\cH,\{\Phi_\mu\},T)$.
Let us demonstrate here that the boundary state corresponding
to the configuration given by (\ref{Dsol1}) and (\ref{Dsol2})
is equivalent to the boundary state of a D$p$-brane ($p$: even).

Inserting the configuration
\beqa
T&=&u\,\sum_{\alpha=0}^p
\wh{p}_\alpha \otimes \gamma^\alpha\\
\Phi^{\alpha} &=& \wh{x}^{\alpha} \otimes {\rm 1}
\;\; (\alpha=0, \cdots, p), \;\;\;\;
\Phi^i=0\;\; (i=p+1, \cdots, 9),
\eeqa
into (\ref{M}), we obtain
\begin{eqnarray}
M_1-M_0^2 &=&
\mat{i\wh x_\alpha P^\alpha-u^2\wh p_\alpha^2,
iu\gamma_\alpha\Pi^\alpha,iu\gamma_\alpha\Pi^\alpha,
i\wh x_\alpha P^\alpha-u^2\wh p_\alpha^2}\\
&=&(i\wh x_\alpha P^\alpha-u^2\wh p_\alpha^2) I
+(iu\Pi^\alpha)\Gamma_\alpha,
\end{eqnarray}
where $\Gamma_\alpha=\left({~~~\gamma_\alpha\atop \gamma_\alpha~~}\right)$
are $SO(p+2)$ gamma matrices.
Then the boundary interaction becomes
\begin{eqnarray}
e^{-S_b(P,\Pi)}&=&
\int[d\eta^\alpha]\,
\Tr_{\cH}{\rm P}
\exp\left\{
\oint d\sigma\left(
\frac{1}{4}\dot\eta^\alpha\eta^\alpha+
i\wh x_\alpha P^\alpha- u^2\wh p_\alpha^2+
iu\Pi^\alpha\eta^\alpha
\right)\right\}\\
&=&
\int[d\eta^\alpha]\,
e^{
\oint d\sigma\left(
\frac{1}{4}\dot\eta^\alpha\eta^\alpha+
iu\Pi^\alpha\eta^\alpha\right)
}
\Tr_{\cH}{\rm P}\,
e^{-\oint d\sigma
H(\wh p,\wh x)},
\label{Bint2}
\end{eqnarray}
where we defined $H(\wh p,\wh x)=u^2\wh p_\alpha^2-i\wh x_\alpha P^\alpha$.
$H(\wh p,\wh x)$ can be thought of as a Hamiltonian of
a point particle with a kinetic term $u^2\wh p_\alpha^2$ and a potential
term $-i\wh x_\alpha P^\alpha$.
We can rewrite (\ref{Bint2}) in terms of the path integral formulation
using the standard formula
\begin{eqnarray}
\Tr_{\cH}{\rm P}\, e^{-\int d\sigma H(\wh p,\wh x)}
=\int[dx] e^{-\int d\sigma L(\dot x,x)}.
\label{key}
\end{eqnarray}
Thus we obtain
\begin{eqnarray}
e^{-S_b(P,\Pi)}&=&
\int[dx^\alpha][d\eta^\alpha]\,
\exp\left\{
\oint d\sigma\left(
\frac{1}{4}\dot\eta^\alpha\eta^\alpha
- \frac{\dot x_\alpha^2}{4u^2}+ix_\alpha P^\alpha
+iu\Pi^\alpha\eta^\alpha
\right)\right\}\\
&=&
\int[dx^\alpha][u d\theta^\alpha]\,
\exp\left\{
\oint d\sigma\left(
-\frac{1}{4u^2}(\dot x_\alpha^2+\theta^\alpha\dot\theta^\alpha)
+ix_\alpha P^\alpha
+i\Pi^\alpha\theta^\alpha
\right)\right\},\nonumber\\
\end{eqnarray}
where $\theta^\alpha= u\eta^\alpha$.

The boundary state of NS-sector is
\begin{eqnarray}
&&e^{-S_b(P,\Pi_\pm)}\ket{D;\pm}_{\rm NS}\\
&=&
\int[dx^\alpha][u d\theta^\alpha]\,
\exp\left\{
\oint d\sigma\left(
-\frac{1}{4u^2}(\dot x_\alpha^2+\theta^\alpha\dot\theta^\alpha)
+ix_\alpha P^\alpha
+i\Pi^\alpha_\pm\theta^\alpha
\right)\right\}
\ket{x=0}\ket{\theta=0;\pm}_{\rm NS}\nonumber\\
\\
&=&
\int[dx^\alpha][u d\theta^\alpha]\,
\exp\left\{
\oint d\sigma\left(
-\frac{1}{4u^2}(\dot x_\alpha^2+\theta^\alpha\dot\theta^\alpha)
\right)\right\}
\ket{x^\alpha, x^i=0}\ket{\theta^\alpha,\theta^i=0;\pm}_{\rm NS}\\
&\rightarrow& \int[dx^\alpha][d\theta^\alpha]\,
\ket{x^\alpha, x^i=0}\ket{\theta^\alpha,\theta^i=0;\pm}_{\rm NS}
~~~(\mbox{as}~u\rightarrow\infty)
\label{uinfty}\\
&=&\ket{Bp;\pm}_{\rm NS}.
\label{uinftyNS}
\end{eqnarray}
In (\ref{uinfty}), we used a $\zeta$-function
regularization trick
\begin{eqnarray}
\prod_{r=1/2,3/2,\dots} u^2 = 1,
\end{eqnarray}
and took a naive limit.
Let us justify (\ref{uinfty}) using oscillator expansion
and $\zeta$-function regularization.
First, we consider the bosonic part.
We adopt the $p=0$ case for simplicity.
\begin{eqnarray}
&&\int[dx]\exp\left(-\oint d\sigma \frac{\dot x^2}{4u^2}\right)\ket{x}\\
&=&\int dx_0\prod_{n=1}^\infty
\left(\int dx_{-n}dx_n\,
e^{-\frac{n}{2u^2} x_{-n} x_n
-\half x_{-n}x_n-a_n^\dag\wt a_n^\dag
+a_n^\dag x_n+x_{-n}\wt a_n^\dag}\right)\ket{x_0}\\
&=&\frac{u\Gamma(u^2)}{\sqrt{2\pi}}\,
e^{-\sum_{n=1}^\infty\left(1-\frac{2}{1+n/u^2}\right)a_n^\dag\wt a_n^\dag}
\ket{0}.
\label{bos}
\end{eqnarray}
We used the $\zeta$-function regularization to obtain (\ref{bos})
(see Appendix \ref{zeta}).

Similarly the fermionic part can also be calculated.
\begin{eqnarray}
&&\int[ud\theta]\exp\left(-\oint d\sigma
\frac{\theta\dot\theta}{4u^2}\right)\ket{\theta;\pm}_{\rm NS}\\
&=&\prod_{r>0}^\infty
\left(\int u^2 d\theta_{-r}d\theta_r\,
e^{-\frac{r}{2u^2} \theta_{-r} \theta_r
-\half\theta_{-r}\theta_r\pm i\psi_r^\dag\wt\psi_r^\dag
+\psi_r^\dag \theta_r\mp
i\theta_{-r}\wt\psi_r^\dag}\right)\ket{0}_{\rm NS}\\
&=&\frac{\sqrt{2\pi}}{\Gamma\left(u^2+\half\,\right)}
e^{\pm \sum_{r>0}^\infty\left(1-\frac{2}{1+r/u^2}\right)
i\psi_r^\dag\wt\psi_r^\dag}
\ket{0}_{\rm NS}.
\label{NSfer}
\end{eqnarray}

Combining (\ref{bos}) and (\ref{NSfer}) together,
we obtain
\begin{eqnarray}
&&e^{-S_b(P,\Pi_\pm)}\ket{D;\pm}_{\rm NS}\\
&=&
\frac{u\Gamma(u^2)}{\Gamma(u^2+\half\,)}\,
e^{-\sum_{n=1}^\infty\left(1-\frac{2}{1+n/u^2}\right)a_n^\dag\wt a_n^\dag
\pm \sum_{r>0}^\infty\left(1-\frac{2}{1+r/u^2}\right)
i\psi_r^\dag\wt\psi_r^\dag}
\ket{0}_{\rm NS}
\label{GamGam}\\
&\rightarrow&
\ket{B0;\pm}_{\rm NS}~~~(\mbox{as}~u\rightarrow\infty),
\end{eqnarray}
which actually agrees with the previous estimation (\ref{uinftyNS}).
This calculation is precisely analogous to
that given in \cite{KrLa,deAl,ArPaSt}.
In fact the coefficient
$\frac{u\Gamma(u^2)}{\Gamma(u^2+1/2)}=
\frac{4^{u^2} \Gamma(u^2)^2}{2\sqrt{\pi}\Gamma(2u^2)}$
in (\ref{GamGam}) is exactly the same as
the factor which plays a crucial role to obtain
the exact D-brane tension in BSFT \cite{KuMaMo2,KrLa,TaTeUe}.

For the RR-sector, note that
\begin{eqnarray}
[ud\theta] &=& u d\theta_0 \prod_{n=1}^\infty u^2 d\theta_{-n}d\theta_n
=d\theta_0 \prod_{n=1}^\infty d\theta_{-n}d\theta_n,
\end{eqnarray}
where we again used the $\zeta$-function trick
\begin{eqnarray}
\prod_{n=1}^\infty u^2 = u^{2\zeta(0)}=u^{-1}.
\end{eqnarray}
Thus the same argument as in (\ref{uinfty})
implies that
\begin{eqnarray}
e^{-S_b(P,\Pi_\pm)}\ket{D;\pm}_{\rm RR}\rightarrow\ket{Bp;\pm}_{\rm RR}
~~~(\mbox{as}~u\rightarrow\infty).
\label{uinftyRR}
\end{eqnarray}

More careful analysis with the $\zeta$-function regularization
as above can also be performed analogously.
\begin{eqnarray}
&&\int[ud\theta]\exp\left(-\oint d\sigma
\frac{\theta\dot\theta}{4u^2} \right)\ket{\theta;\pm}_{\rm RR}\\
&=&\prod_{n=1}^\infty
\left(\int u^2 d\theta_{-n}d\theta_n\,
e^{-\frac{n}{2u^2} \theta_{-n} \theta_n
-\half\theta_{-n}\theta_n\pm i\psi_n^\dag\wt\psi_n^\dag
+\psi_n^\dag \theta_n\mp
i\theta_{-n}\wt\psi_n^\dag}\right)
 \int u d\theta_0\, e^{i\Pi_{0\pm}\theta_0} \ket{D;\pm}^{(0)}_{\rm RR}
\nonumber\\ \\
&=&\frac{\sqrt{2\pi}}{u \Gamma(u^2)}\,
e^{\pm \sum_{n=1}^\infty\left(1-\frac{2}{1+n/u^2}\right)
i\psi_n^\dag\wt\psi_n^\dag}
\ket{B0;\pm}^{(0)}_{\rm RR},
\label{RRfer}
\end{eqnarray}
where we used the relation (\ref{zero}) for the ground state.
The coefficient exactly cancels that in bosonic part (\ref{bos}),
as expected from the quantization of the RR-charge,
and we obtain
\begin{eqnarray}
&&e^{-S_b(P,\Pi_\pm)}\ket{D;\pm}_{\rm RR}\\
&=&
e^{-\sum_{n=1}^\infty\left(1-\frac{2}{1+n/u^2}\right)a_n^\dag\wt a_n^\dag
\pm\sum_{n=1}^\infty
\left(1-\frac{2}{1+n/u^2}\right)
i\psi_n^\dag\wt\psi_n^\dag}
\ket{B0;\pm}^{(0)}_{\rm RR}\\
&\rightarrow&
\ket{B0;\pm}_{\rm RR}~~~(\mbox{as}~u\rightarrow\infty).
\end{eqnarray}

Finally, (\ref{uinftyNS}) and (\ref{uinftyRR}) imply that
 the boundary state of non-BPS D-instantons
with the boundary interactions (\ref{Bint}) is exactly
equal to the boundary state of BPS D$p$-brane (\ref{BPS})
in the limit $u\rightarrow\infty$,
\begin{eqnarray}
\ket{\wh D(-1)}_{S_b}
\rightarrow \ket{Dp} ~~~(\mbox{as}~u\rightarrow\infty),
\end{eqnarray}
as promised.

What can we learn from this example?
The boundary states we reviewed in the last subsection,
 such as (\ref{Bp}),
are constructed from the geometric information of the D-branes.
Namely, we first
decided how the world-volume of the D-brane is
embedded in the space-time, and
arranged a suitable coherent state to construct the
D-brane boundary state.
On the other hand, as we have seen,
D-brane configurations are represented by
analytic data $(\cH,\{\Phi_\mu\},T)$ , i.e. the spectral triple,
in the K-matrix theory.
We can construct the D-brane boundary state corresponding
to each triple $(\cH,\{\Phi_\mu\},T)$ as given in (\ref{bdry2})
and (\ref{Bint}). In this subsection, we have shown,
using an explicit example, that these two constructions
are actually equivalent. This is the stringy
realization of the isomorphism (\ref{isom})
between topological K-homology and analytic one.
The key relation is (\ref{key}), which translates
 the analytic data (in the operator formalism)
into the geometric one (in the path integral).
It is interesting to note that
the Hilbert space $\cH$, which is interpreted as the space of
Chan-Paton indices of non-BPS D-instantons, is translated
into the Hilbert space of the quantum mechanics
of the boundary degrees of freedom defined by the path integral
in the right hand side of (\ref{key}).
The analogous statement
has been observed in noncommutative geometry
\cite{Is,Ok}, but our analysis shows that this correspondence
is also realized even in the commutative cases.

\section{Conclusions and Discussions}
\label{Discuss}

In this paper, we studied the matrix theory based on
non-BPS D-instantons in type IIA string theory and
D-instanton - anti D-instanton system in type IIB string theory,
which we called K-matrix theory.
The configurations with finite action
are identified with spectral triples,
and the geometry represented by the spectral triples
are interpreted as the geometry on the world-volume of
higher dimensional D-branes.
Furthermore, we claimed
that the configurations of D-branes in the K-matrix theory
are classified by K-homology.
We also constructed the boundary states corresponding
to the configurations of the K-matrix theory,
and explicitly showed that the canonical triples
represent higher dimensional D-branes.

It would be interesting to investigate the
relation between
 our proposal that D-branes are
represented by the spectral triples
and the description of
D-branes as objects of
the derived category of coherent sheaves
\cite{Sh,Do}.
Actually, they are closely related.
In fact, an element of K-homology group in algebraic geometry
can be obtained by an object of
the derived category of coherent sheaves \cite{Sh}.
Here, K-homology group in algebraic geometry,
denoted $K_0'(X)$, is defined as the
Grothendieck group of coherent algebraic sheaves on
algebraic variety  $X$.
One can define a natural map \cite{BaDo}
\begin{eqnarray}
\alpha: K_0'(X)\rightarrow K_0^{top}(X),
\end{eqnarray}
though this is {\it not} isomorphic in general.
The reason that $\alpha$ fails to be isomorphic can
be understood from the fact that
$K^{top}_0(X)$ does not respect holomorphic structure,
while $K'_0(X)$ does.
See \cite{BaDo,Sh} for more detail.

There are many important issues, which we left for
the future study. First of all,
we have not made an argument about the
consistency of the theory as a quantum theory.
Since the variables in the theory are operators acting
on an infinite dimensional Hilbert space,
it is not clear that all the physical quantities
remain finite. The action of the theory
should be determined, to a certain extent,
 by the consistency of the theory.
Similarly, we were not careful about the
choice of the space-time manifold $X$.
Since a consistent background
should be a solution of the equations of motion of
supergravity, there should be some restrictions
for the choice.
Some related arguments about the formulation of the
general matrix theories in curved backgrounds can be
found in \cite{Do2,Do3,DoKaOo}.

In addition,
in section \ref{embed} and section \ref{K-homo},
 we chose the space-time algebra $\cA$
independent of the choice of the background,
in which the K-matrix theory is supposed to be formulated.
We do not know the precise relation between the closed string background
and the space-time algebra we can choose.
In particular,
there are many D-brane configurations
that are not included in the space-time represented by the fixed algebra.
For example, we do have noncommutative D-brane configurations,
even if the background is commutative.
In the classification of stable D-brane configurations,
we classified the D-brane configurations embedded in a fixed
space-time manifold.
There might be a possibility that the D-branes decay
through going `outside' the space-time that we fixed.

The appearance of the closed strings
in the K-matrix theory is also a very interesting subject.
If the theory consistently formulate the
type II string theory, there should be closed strings.
Unfortunately, K-homology is not powerful enough to
classify the fundamental strings and NS5-branes,
and we failed to incorporate these objects into the classification.
See \cite{KaKu} for the recent related work.

One of the interesting features of the K-matrix theory
in contrast to other matrix theories is that
we can construct arbitrary numbers of the D-branes.
Even the configuration with `nothing'
 is also included as a configuration of the
theory.
This fact may have interesting applications to
the formulation of M-theory.
In the BFSS matrix theory \cite{BFSS},
 the number $N$ of the D-particles
are fixed to a finite or infinite value.
Therefore,
it can only represent M-theory
with fixed momentum along the light-like or eleventh direction.
On the other hand, in the K-matrix theory (based on infinite
number of either non-BPS D-instantons or D0-\AD0 pairs),
there are no such restrictions,
and it is quite easy to construct a configuration
with arbitrary numbers of D-particles
and anti D-particles.
Therefore the K-matrix theory could
provide a much wider framework to study M-theory.

There is another intriguing structure of the
IIA K-matrix theory based on non-BPS D-instantons,
 which is the same structure observed
in \cite{Ho} for the world-volume theory
of non-BPS D9-branes.
Recall that the bosonic part of the K-matrix theory
consists of ten scalars $\Phi^\mu$ $(\mu=0,1,\dots,9)$
and one tachyon $T$,
which transforms as  $10\oplus 1$ representation
under the ten dimensional Lorentz transformation.
The fermions in the theory are $\chi_L$ and $\chi_R$,
which belong to left and right handed spinor
representations of the Lorentz group, respectively.
They coincide with the
ten dimensional decomposition of
the vector and spinor representations of
eleven dimensional Lorentz group!
Of course, it is hard to consider the tachyon as
the scalar $\Phi^{10}$
which corresponds to the eleventh direction,
since eleven dimensional Lorentz symmetry is
broken explicitly.
It would be interesting,
if this fact should be a clue to
a formulation of M-theory with explicit
eleven dimensional Lorentz symmetry.

Although there is a possibility that the fields $\Phi^\mu$ and $T$ are
sufficient to describe the whole things,
the precise action of the K-matrix theory may be given
by the boundary or cubic string field theory of the infinitely many
non-BPS D-instantons.
In such cases, infinitely many matrices or operators
should be considered, and
we might have to generalize
 the choice of the triples $(\cH,\wh\cA,T)$, which represent
the configurations.

We hope to come back to these problems in later publication.

%
%

\vskip6mm\noindent
{\bf Acknowledgements}

\vskip2mm
We would like to thank
K. Hosomichi,
H. Moriyoshi, T. Takayanagi, Y. Terashima and T. Uesugi
for useful discussions.
This work was supported in part by JSPS Research Fellowships for Young
Scientists.
We appreciate the organizers of
Summer Institute 2000 and 2001 at Yamanashi,
where a part of this work was done.
S.S. is also grateful to the theoretical physics group
at University of Tokyo for hospitality, and the NSF (\# 9724831) for
making possible the collaboration between University of Tokyo and
USC.

\appendix
\setcounter{equation}{0}
\section{Chern-Simons term of $N$ non-BPS D-instantons}
\label{App1}

In this Appendix \ref{App1}, we prove the topological invariance of CS-term
of $N$ non-BPS D-instantons, namely the invariance of the CS-term under
$C\rightarrow C+d\Lambda$.
First we define
\beq
C(x)=\int dk e^{ikx} C(k),
\eeq
and
\beq
J(k)= \Tr_2 \left( \sigma^1  \Tr_N \left(
e^{ik \Phi+ Z^2} \right) \right),
\eeq
where
\beqa
iZ &=& -i \Phi^\mu \psi_2^\mu+ T \sigma^1, \\
C(x) &=& \sum_{n:odd} C_{\mu_1\cdots\mu_n}(x)
\, \psi_1^{\mu_1} \cdots \psi_1^{\mu_n},
\eeqa
as defined in (\ref{Z}), (\ref{C}). Note that $n$ is odd here.

Then we can rewrite the CS-term of $N$ non-BPS D-instantons
(\ref{CSterm}) as
\beq
S_{CS}= \Tr_{\psi} \left(
\int dk\,C(k) J(k)\right).
\eeq
In order to show that
the CS-term is invariant under the gauge transformation
$C\rightarrow C+d\Lambda$,
it is sufficient to check that the CS-term vanishes if we
take $C(k)=k_l \Lambda(k) \psi_1^{l}$.
where $\Lambda(k)= \Lambda_{i_1 \cdots i_{n-1}}
\psi_1^{i_1} \cdots \psi_1 ^{i_{n-1}}$.
This condition is indeed satisfied as
\beqa
S_{CS} &=& \Tr_\psi \left(
\int dk \Lambda(k) k_l \psi^l_1 J(k) \right)
= \Tr_\psi \left(
\int dk \Lambda(k) \frac{1}{2} \{ k_l \psi^l_1, J(k)\} \right)
\CR
&=&-\Tr_2 \left( \sigma^1  \Tr_\psi \left(
\int dk \Lambda(k) \underbar{\mbox{Sym}}\Tr_N \left(
[k \Phi, Z] e^{ik \Phi +Z^2}
\right) \right) \right)
\CR
&=&i \Tr_2 \left( \sigma^1  \Tr_\psi \left(
\int dk \Lambda(k) \Tr_N \left(
[ e^{ik \Phi +Z^2},Z]
\right) \right) \right)=0,
\eeqa
where
$\underbar{\mbox{Sym}}$ means symmetrization w.r.t.
$[k \Phi, Z]$ and $(ik \Phi +Z^2)$.
Thus we have confirmed the invariance
 of the CS-term under the gauge transformation of the RR-fields.
Though we demonstrated the calculation for
the CS-term of $N$ non-BPS D-instanton,
it is straightforward to generalize to other systems.
In particular, this proof is also applicable for the CS-terms
for the BPS D-branes, which has been shown in \cite{OkOo}.
It provides a more simple and general proof for the invariance
of CS-term under the gauge transformation.


Since the CS-term is a linear functional of $C(x)$,
it should be written as $S_{CS}=\int_X C(x) I(x)$,
where $I(x)=I^1(x)+I^2(x)+\cdots$ and $I^i$ is a $i$-form
determined from $\Phi$ and $T$ through (\ref{CSterm}).\footnote{
Here we assume a smoothness of the configuration.}
Using this form, the invariance under $C\rightarrow C+d\Lambda$
implies that $I(x)$ is closed form, i.e.
it defines a cohomology class of $X$.
Therefore we find that
any configuration of $\Phi$ and $T$
determines a cohomology class which is nothing but the
RR-charge of the configuration.

%

\section{$\zeta$-function regularization}
\label{zeta}

We summarize the zeta-function regularization formulae
used in section \ref{BdryState}. (See also \cite{KrLa}.)
\begin{eqnarray}
\prod_{r=1/2,3/2,\cdots} A
 = \exp\left(\log A \sum_{n=1}^\infty (n+1/2)^{-s}\right)\Bigg|_{s=0}
 = A^{\zeta(0,1/2)} = 1
\end{eqnarray}
\begin{eqnarray}
\prod_{n=1}^\infty A
 = \exp\left(\log A \sum_{n=1}^\infty n^{-s}\right)\Bigg|_{s=0}
 = A^{\zeta(0)} = A^{-1/2}
\end{eqnarray}
\begin{eqnarray}
\prod_{n=1}^\infty(n+a)^{-1}
&=&\exp\left(\frac{d}{ds} \sum_{n=1}^\infty(n+a)^{-s}\right)\Bigg|_{s=0}\\
&=& \exp\left(
\frac{d}{ds}\left(\zeta(s,a)-a^{-s}\right)\right)\Bigg|_{s=0}
= \frac{\Gamma(a+1)}{\sqrt{2\pi}}
\end{eqnarray}
\begin{eqnarray}
\prod_{r=1/2,3/2,\cdots} r= \frac{\sqrt{2\pi}}{\Gamma(1/2)}=\sqrt{2}
\end{eqnarray}



\end{document}